\documentclass[aps,pra,twocolumn,showpacs,superscriptaddress,floatfix,nofootinbib]{revtex4-1}

\usepackage{graphicx}
\usepackage{amssymb}
\usepackage{natbib}
\usepackage{amsmath}
\usepackage{etoolbox}
\usepackage{color}

\begin{document}

\title{Modeling sympathetic cooling of molecules by ultracold atoms}

\author{Jongseok Lim}
\affiliation{Centre for Cold Matter, Blackett Laboratory, Imperial College London,
Prince Consort Road, London SW7 2AZ, United Kingdom}
\author{Matthew D. Frye}
\affiliation{Joint Quantum Centre (JQC) Durham-Newcastle, Department of
Chemistry, Durham University, South Road, Durham DH1 3LE, United Kingdom}
\author{Jeremy M. Hutson}
\affiliation{Joint Quantum Centre (JQC) Durham-Newcastle, Department of
Chemistry, Durham University, South Road, Durham DH1 3LE, United Kingdom}
\author{M. R. Tarbutt}
\email{m.tarbutt@imperial.ac.uk}
\affiliation{Centre for Cold Matter, Blackett Laboratory, Imperial College London,
Prince Consort Road, London SW7 2AZ, United Kingdom}

\begin{abstract}
We model sympathetic cooling of ground-state CaF molecules by ultracold Li and
Rb atoms. The molecules are moving in a microwave trap, while the atoms are
trapped magnetically. We calculate the differential elastic cross sections for
CaF-Li and CaF-Rb collisions, using model Lennard-Jones potentials adjusted to
give typical values for the s-wave scattering length. Together with trajectory
calculations, these differential cross sections are used to simulate the
cooling of the molecules, the heating of the atoms, and the loss of atoms from
the trap. We show that a hard-sphere collision model based on an
energy-dependent momentum transport cross section accurately predicts the
molecule cooling rate but underestimates the rates of atom heating and loss.
Our simulations suggest that Rb is a more effective coolant than Li
for ground-state molecules, and that the cooling dynamics are
less sensitive to the exact value of the s-wave scattering length when Rb is
used. Using realistic experimental parameters, we find that molecules can be
sympathetically cooled to 100\,$\mu$K in about 10\,s. By applying evaporative
cooling to the atoms, the cooling rate can be increased and the final
temperature of the molecules can be reduced to 1\,$\mu$K within 30\,s.
\end{abstract}
\date{\today}
\maketitle

\section{Introduction}

Ultracold molecules are important for several applications in physics and
chemistry. Cold molecules have already been used to test theories that extend
the Standard Model of particle physics, for example by measuring the electron's
electric dipole moment \cite{Hudson(1)11, Baron(1)14} or searching for changes
in the fundamental constants \cite{Hudson:2006, Truppe(1)13}. The precision of
those measurements can be improved by cooling the molecules to far lower
temperatures \cite{Tarbutt(1)09, Tarbutt(1)13}. A lattice of ultracold polar
molecules makes a well-controlled many-body quantum system where each particle
interacts with all others through the long-range dipole-dipole interaction.
This array can be used as a model system to study other strongly-interacting
many-body quantum systems whose complexity is far too high to simulate on a
computer \cite{Micheli:2006}. Ultracold polar molecules offer several
advantages for storing and processing quantum information \cite{DeMille:2002,
Andre:2006}, notably strong coupling to microwave photons and, through
dipole-dipole interactions, to one another. The availability of ultracold
molecules will also open up opportunities for studying and controlling chemical
reaction dynamics in a whole new regime \cite{Krems:PCCP:2008}.

Some species of ultracold polar molecules can be produced by association of
ultracold atoms, either by photoassociation \cite{Deiglmayr(1)08,
Shimasaki(1)15} or by magnetoassociation through a Feshbach resonance
\cite{Ni(1)08, Takekoshi:RbCs:2014, Molony:RbCs:2014}. Often though, the
molecules of interest cannot be formed this way, and then more direct cooling
methods are needed. Molecules have been magnetically trapped at temperatures of
about 0.5\,K by buffer-gas cooling with cryogenic helium
\cite{Weinstein:CaH:1998, Tsikata(1)10}. Molecules in supersonic beams have
been decelerated to rest and then trapped electrically and magnetically,
typically with temperatures in the range 1-50\,mK \cite{Bethlem:trap:2000,
Sawyer:2007}. Recently, laser cooling has been applied to SrF
\cite{Shuman:2010, Barry:beam:2012}, YO \cite{Hummon:2013} and CaF
\cite{Zhelyazkova(1)14}, and a magneto-optical trap of SrF has been
demonstrated, producing molecules at a temperature of a few mK and a density of
4000\,cm$^{-3}$ \cite{Barry(1)14, McCarron(1)15}. It is likely that higher
densities will be reached using more efficient loading methods, and lower
temperatures may be reached if sub-Doppler cooling mechanisms are effective.

A promising method to cool molecules to lower temperatures is sympathetic
cooling through collisions with ultracold atoms. The main difficulty with this
approach is that static electric and magnetic traps can only confine molecules
in weak-field seeking states, but the lowest-energy state is always
high-field-seeking. It follows that inelastic collisions can heat the molecules
or can transfer them from trapped to untrapped states. This observation has
motivated experimental \cite{Parazzoli:2011} and theoretical work
\cite{Parazzoli:2011, Lara:PRA:2007, Wallis:MgNH:2009,
Gonzalez-Martinez:hyperfine:2011, Wallis:LiNH:2011, Tscherbul:poly:2011,
Gonzalez-Martinez:H+mol:2013} to search for atom-molecule systems where the
ratio of elastic to inelastic collision cross sections is large. However, for
most systems of interest, this ratio is too small for sympathetic cooling to
work well. Notable exceptions are the Mg + NH system \cite{Wallis:MgNH:2009},
and the use of ultracold hydrogen as a coolant
\cite{Gonzalez-Martinez:H+mol:2013}, but the experimental realization of those
systems is exceptionally challenging. An alternative approach is to use a
dynamic trap, which could be an alternating current (ac) trap, an optical
dipole trap, or a microwave trap, so that molecules can be confined in their
lowest energy states. In this case, inelastic collisions can only excite the
molecule, but the energy available in the collision is typically too small for
that and so all inelastic channels are energetically inaccessible.

In previous work \cite{Tokunaga:2011} sympathetic cooling of a cloud of LiH
molecules by ultracold Li atoms was simulated using a very simple model. The scattering was assumed to be isotropic, corresponding to either s-wave
scattering or classical collisions of hard spheres. This is appropriate for collisions at
very low energy. However, the differential cross sections at higher collision
energies are typically peaked at low deflection angles, because many collisions
sample mainly the long-range attraction. In the present work, we introduce a
new collision model that takes account of the full energy dependence of the differential cross sections. We
show that this model produces significantly slower sympathetic cooling in the
early stages than the original energy-independent hard-sphere model. We also
consider approximations to the full model and show that a model that uses
hard-sphere scattering based on the energy-dependent transport cross section
$\sigma_\eta^{(1)}$ \cite{Frye:2014} produces accurate results for the cooling
of the molecules but not for heating and loss of the coolant atoms.

The previous modeling work \cite{Tokunaga:2011} explored sympathetic cooling in
three different types of trap: a static electric trap, an alternating current
(ac) trap, and a microwave trap. A static electric trap can confine molecules
only in rotationally excited states, and it was found that for Li+LiH the ratio
of elastic to rotationally inelastic collisions was too small for such
molecules to be cooled before they were ejected from the trap. An ac trap can
confine molecules in the rotational ground state, so there are no inelastic
collisions, but elastic collisions can transfer molecules from stable to
unstable trajectories and it was found that this eventually causes all the
molecules to be lost. A microwave trap \cite{DeMille:2004, Dunseith(1)15} can
confine molecules in the absolute ground state, around the antinodes of a
standing-wave microwave field, and sympathetic cooling in this trap was found
to be feasible on a timescale of 10\,s \cite{Tokunaga:2011}. The microwave trap
brings the benefits of a high trap depth and large trapping volume for polar
molecules, especially compared to an optical dipole trap. In the present work,
we simulate sympathetic cooling in a microwave trap in detail. We consider the
following specific, experimentally realistic, scenario. Cold CaF molecules are
produced either in a magneto-optical trap \cite{Barry(1)14, McCarron(1)15} or
by Stark deceleration \cite{Wall(1)11, VanDenBerg(1)14}. In the first case the
temperature might be about 2\,mK, and in the second about 30\,mK. The molecules
are loaded into a magnetic trap, and then transported into a microwave trap.
Here, the molecule cloud is compressed in order to improve the overlap with the
atomic coolant, and this raises the initial temperature of the molecules to
20\,mK and 70\,mK respectively. A distribution of atoms, either $^{7}$Li or
$^{87}$Rb, with an initial temperature of 100\,$\mu$K, is trapped magnetically
and is overlapped with the cloud of molecules. We simulate the way in which
elastic collisions reduce the molecular temperature towards the atomic
temperature. Black-body heating out of the rovibrational ground state can be
reduced below $10^{-4}$\,s$^{-1}$ by cooling the microwave trap to 77\,K
\cite{Buhmann(1)08}.

We start by describing our scattering calculations and the cross sections we
obtain. Then we describe the simulation method we use, and study how the choice
of collision model affects the simulation results. Next, we examine the cooling
dynamics and evaluate which coolant, Rb or Li, is likely to be the best in
practical situations. Because the cross section is very sensitive to the exact
form of the atom-molecule interaction potential, especially at low energies, we
study sympathetic cooling for a range of typical values of the s-wave
scattering length. In addition to cooling the molecules, collisions either heat
the atoms, raising the final temperature, or eject atoms from the trap,
reducing the atomic density. These effects are particularly important if the
atom number does not greatly exceed the molecule number. We study these effects
and explain the results in terms of appropriate partial integrals over
differential cross sections. Finally, we investigate how evaporative cooling of
the atoms can be used to speed up the sympathetic cooling rate and lower the
final temperature obtained.

\section{Scattering calculations}
\label{sec:crosssection}

\begin{figure*}[tb]
 \centering
 \includegraphics[width=0.75\textwidth]{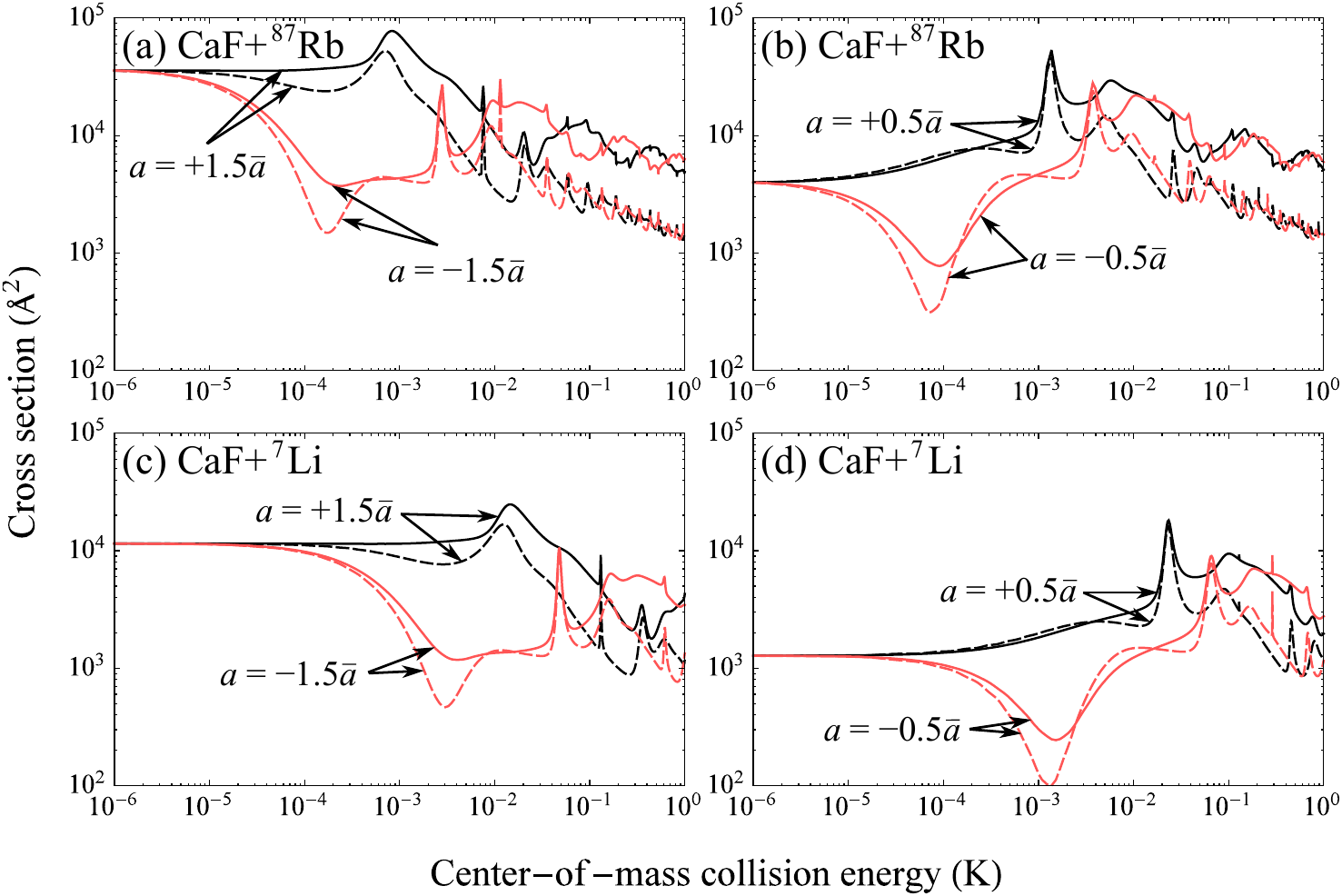}
\caption{\label{crosssection} (Color online) Total elastic cross section,
$\sigma_{\text{el}}$ (solid lines), and transport cross section,
$\sigma_{\eta}^{(1)}$ (dashed lines), for positive (black) and negative
(red/gray) signs of the scattering length. (a) CaF+$^{87}$Rb,
$\left|a\right|=1.5\bar{a}$; (b) CaF+$^{87}$Rb, $\left|a\right|=0.5\bar{a}$;
(c) CaF+$^{7}$Li, $\left|a\right|=1.5\bar{a}$; (d) CaF+$^{7}$Li,
$\left|a\right|=0.5\bar{a}$.}
\end{figure*}

Exact scattering calculations on systems as complex as Li+CaF and Rb+CaF are
not currently feasible. The combination of a deep chemical well, very large
anisotropy of the interaction potential, and small CaF rotational constant mean
that a very large rotational basis set would be needed for convergence. In
addition, even if converged results could be achieved, uncertainties in the
potential surface mean that no single calculation could be taken to represent
the true system and many calculations on many surfaces would be needed to
explore the range of possible behaviors \cite{Cvitas:li3:2007}. Instead we
model the interactions with a simple single-channel model potential which we
choose to be the Lennard-Jones potential, $V(r)=-C_6/r^6+C_{12}/r^{12}$, where
$r$ is the intermolecular distance. We have shown previously \cite{Frye:2015}
that, while a simple single-channel model cannot be expected to reproduce a
full coupled-channel calculation, it can quantitatively reproduce the {\em
range} of behaviors shown by full calculations.

We obtain Lennard-Jones parameters for Li+CaF from {\em ab initio} calculations
\cite{Morita:unpub:2015}. We obtain $C_{6 \rm ,Li+CaF} =1767\,E_{\rm h}a_0^6$
from direct fitting to the isotropic part of the long-range potential, where
$E_{\rm h}$ is the Hartree energy and $a_{0}$ is the Bohr radius. We set $C_{12
\rm ,Li+CaF}=2.37 \times 10^{7}\,E_{\rm h}a_0^{12}$ to reproduce the depth of
the complete potential, which is 7224 cm$^{-1}$. We use the depth of the
complete potential in preference to the depth of the isotropic part of the
potential because the very large anisotropy at short-range means the isotropic
part of the potential is not representative of the interaction. To obtain a
$C_6$ parameter for Rb +CaF we first separate $C_{6 \rm ,Li+CaF}$ into
induction and dispersion contributions. Induction contributions for both
systems are readily calculated from known values of the CaF dipole moment
\cite{Childs:1984} and the static polarizabilities of the atoms
\cite{Derevianko:2010}. The dispersion contribution for Rb+CaF can then be
calculated from the dispersion contribution for Li+CaF using Tang's combining
rule \cite{Tang:1969} with known homonuclear diatomic dispersion coefficients
\cite{Derevianko:2010}, atomic polarizabilities \cite{Derevianko:2010} and a
calculated CaF polarizability of $\alpha_{\rm CaF}=137 a_0^3$. The sum of these
contributions gives $C_{6 \rm ,Rb+CaF}=3084\,E_{\rm h}a_0^6$. We estimate, by
analogy to calculations on methyl fluoride \cite{Lutz:2014}, that the well
depth for Rb+CaF will be about 2.5 times shallower than for Li+CaF. This sets
$C_{12 \rm ,Rb+CaF}=1.8 \times 10^{8} \,E_{\rm h}a_0^{12}$.

For our purposes, the key property of a potential is the s-wave scattering
length, $a$, that it produces. In the present work, we vary the $C_{12}$
coefficient over a small range (with $C_6$ fixed) to vary the scattering
length. We focus on four typical scattering lengths, $a= -1.5\bar{a}$,
$-0.5\bar{a}$, $+0.5\bar{a}$, $+1.5\bar{a}$, where $\bar{a}$ is the mean
scattering length of Gribakin and Flambaum \cite{Gribakin:1993}, $\bar{a}=20.2$
\AA\ for Li+CaF and $35.7$ \AA\ for Rb+CaF.

Discussions of thermalization have usually assumed that the relevant cross
section is the elastic cross section $\sigma_{\rm el}$, which is the unweighted
integral of the differential cross section $d\sigma/d\omega$,
\begin{equation}
\sigma_{\rm el} = 2\pi \int \frac{d\sigma}{d\omega}\sin\Theta\,d\Theta,
\end{equation}
where $d\omega$ is an element of solid angle and $\Theta$ is the deflection
angle in the center-of-mass frame. However, small-angle scattering contributes
fully to $\sigma_{\rm el}$ but contributes relatively little to thermalization.
The transport cross section that takes proper account of this is
$\sigma_\eta^{(1)}$ \cite{Frye:2014},
\begin{equation}
\sigma_\eta^{(1)}= 2\pi \int \frac{d\sigma}{d\omega} (1-\cos\Theta)\sin\Theta\,d\Theta.
\label{eqn:sig_eta}
\end{equation}

In the present work, scattering calculations are carried out using the MOLSCAT
package \cite{molscat:v14}. We use the DCS post-processor \cite{DCS} to
calculate differential cross sections, and the SBE post-processor \cite{SBE} to
calculate $\sigma_\eta^{(1)}$.

The calculated elastic and transport cross sections for Li+CaF and Rb+CaF are
shown in Fig.\,\ref{crosssection} for a variety of scattering lengths. At low
energy, in the s-wave regime, the cross sections have constant limiting values
of $4 \pi |a|^2$. This is the same for both $\sigma_{\rm el}$ and
$\sigma_\eta^{(1)}$, because pure s-wave scattering is isotropic. The cross
sections for positive and negative scattering lengths go to the same low-energy
limit. However, as energy increases, the cross sections all diverge from one
another. Those for negative scattering lengths, especially $a=-0.5\bar{a}$, show
dramatic Ramsauer-Townsend minima as the scattering phase
shift, and hence the s-wave cross section, passes through a zero
\cite{Child:1974}. For $\sigma_\eta^{(1)}$ this minimum is further deepened by
destructive interference between s-wave and p-wave scattering \cite{Frye:2014}.
For $a=+1.5 \bar{a}$ a peak in both cross sections is seen (near
$~10^{-3}$ K for Rb+CaF). This is a d-wave feature corresponding to the energy
of the centrifugal barrier maximum. At higher energies, there are various shape
resonances present for all cases. Nevertheless, once many partial waves
contribute, the cross sections become less dependent on scattering length and
approach classical limits.

It may be noted that the cross sections for the two systems for the same value
of $a/\bar{a}$ are very similar, apart from constant factors in energy
and cross section. In fact they would be nearly identical if the cross sections
were in units of $\bar{a}^2$ and energy in units of
$\bar{E}=\hbar^2/(2\mu\bar{a}^2)$ \cite{Frye:2014, Gao:2001}, where
$\bar{E}=9.51$ mK for Li+CaF and $0.543$ mK for Rb+CaF. This scaling means
that, while the Rb+CaF cross sections are almost independent of scattering
length at 10 mK and above, the Li+CaF cross sections are highly sensitive to
scattering length at any energy below 100 mK.

For stationary atoms the molecular kinetic energy in the laboratory frame,
$E^\text{lab}_\text{CaF}$, is related to the collision energy in the
center-of-mass frame, $E^\text{CM}$, by
$E^\text{lab}_\text{CaF}=(m_{\text{CaF}}/\mu)E^\text{CM}$, where $\mu =
m_{\text{CaF}}m_{\text{at}}/(m_{\text{CaF}}+m_{\text{at}})$ is the reduced mass
of the collision system, $m_{\text{CaF}}$ is the molecular mass and
$m_{\text{at}}$ is the atom mass. The ratio
$E^\text{lab}_\text{CaF}/E^\text{CM}$ is 9.40 for Li+CaF and 1.68 for Rb+CaF.
This introduces a further energy scaling between the two systems in addition to
the difference in $\bar{E}$.

Because the molecules are in the ground state, and the rotational excitation
energy is far greater than the available collision energy, we assume that there
are no inelastic collisions. It is known that there can be molecule-molecule
inelastic collisions in the presence of the microwave field, even when the
microwave frequency is well below the first rotational resonance
\cite{Kajita(1)07, Avdeenkov:2009}. This is a concern for evaporative cooling
of molecules, but less so for sympathetic cooling, where the density of
molecules can be low. It is worth studying whether there can be atom-molecule
inelastic collisions induced by the microwave field, but that is beyond the
scope of this paper.

\section{Simulation method}
\label{sec:method}

We assume that ground state CaF molecules are confined around the central
antinode of a standing-wave microwave field, formed at the center of an open
microwave cavity \cite{Dunseith(1)15}. The interaction potential of the
molecules with the microwave field is
\begin{equation}
U({\bf r}) = -\Delta U \exp\left[-\frac{x^{2}}{w_x^{2}}-\frac{y^{2}}{w_y^{2}}\right]
\cos^{2}\left(\frac{2\pi z}{\lambda}\right),
\end{equation}
where $\Delta U$ is the trap depth and we take $\Delta U/k_{\text{B}}=400$\,mK,
$w_x=16.3$\,mm, $w_y=15.3$\,mm, and $\lambda=21.3$\,mm \cite{Dunseith(1)15}.
The initial phase-space distribution of the molecules is assumed to be
\begin{eqnarray}
\label{phasespacedensity}
f({\bf r},{\bf p}) &=&\frac{n_{0,\text{CaF}}}{(2 \pi m_{\rm CaF}
k_{\text{B}} T)^{3/2}}\nonumber\\
&\times& \exp\left[-\frac{U({\bf r})-U(0)\! +
p^2/(2m_{\rm CaF})}{k_{\text{B}} T}\right],
\end{eqnarray}
where $T$ is the initial temperature of the molecules and $n_{0,\text{CaF}}$ is
the initial density at the center of the trap, which is fixed such that the
total number of molecules is $N_{\text{CaF}}=10^{5}$. For most simulations, we
take $T=70$\,mK in order to study sympathetic cooling from a high temperature.
A distribution of ultracold atoms is overlapped with the molecules. The atoms
are in a harmonic magnetic trap whose depth is 1\,mK. We assume that the
distribution of atoms in phase space depends only on their energy. Therefore,
at all times, the atoms have a Gaussian spatial distribution and a thermal
velocity distribution with temperature $T_{\text{at}}$. They have an initial
temperature of 100\,$\mu$K, an initial central density of $10^{11}$\,cm$^{-3}$,
and an initial number of $10^{9}$. The corresponding initial $1/e$ radius is
1.2\,mm. This initial temperature and density can be reached by first
collecting and cooling the atoms in a magneto-optical trap, followed by a brief
period of sub-Doppler cooling in a molasses before loading into the magnetic
trap. For Rb, polarization gradient cooling is an effective sub-Doppler cooling
mechanism, while for Li velocity-selective coherent population trapping in a
gray molasses can be used \cite{Grier(1)13, Burchianti(1)14}. Our approximation
that the molecules are confined only by the microwave field, and the atoms only
by the magnetic field, is a reasonable one, though our model could be extended
to use the complete potential of both species in the combined fields.

For each molecule, the simulation proceeds as follows. We solve the equation of
motion in the microwave trap for a time step $\Delta t$ which is much smaller
than the mean time between collisions. Then, using the current position, ${\bf
r}$ and velocity, ${\bf v}$, of the molecule, we determine whether or not there
should be a collision as follows. The velocity of an atom is chosen at random
from a thermal distribution with temperature $T_{\text{at}}$. From the atomic
and molecular velocities we calculate the collision energy in the
center-of-mass frame, $E^\text{CM}$. The collision probability is
$P=n_{\text{at}}({\bf r}) \sigma(E^\text{CM}) v_{\text{r}} \Delta t$, where
$v_{\text{r}}$ is the relative speed of the atom and molecule, $n_{\text{at}}$
is the atomic density, and $\sigma(E^\text{CM})$ is either $\sigma_{\text{el}}$
or $\sigma_{\eta}^{(1)}$ (see Sec.\,\ref{Sec:CollisionModels}). A random number
is generated in the interval from 0 to 1, and if this is less than $P$ a
collision occurs. If there is no collision, the velocity of the molecule is
unchanged. If there is a collision, the velocities are transformed into the
center-of-mass frame, a deflection angle is determined as described below, and
the new velocities transformed back into the laboratory frame. If the new total
energy (kinetic energy plus trapping potential) is sufficient for
the atom to escape from the trap, the atom, and its energy prior to the
collision, are removed. The change in energy is shared among all the remaining
atoms. Otherwise, the atom remains in the trap and the change in kinetic energy
is shared between all the atoms. This algorithm is followed for each molecule
in the distribution. The density and temperature of the atom cloud are updated
to account for the atom loss and atom heating at this time step, and then the
simulation proceeds to the next time step.

With our choice of trap depth and initial atom temperature, there is a small
evaporative cooling effect due to atom-atom collisions. For Rb, over the 50\,s
timescale of our simulations, 8\% of the atoms are lost and the temperature
falls to $80\,\mu$K. Prior to Sec.\,\ref{sec:evaporative}, we neglect this
evaporative cooling effect in our simulations because we wish to isolate
effects that are due to atom-molecule collisions. Then, in
Sec.\,\ref{sec:evaporative}, we include atom-atom collisions and explore the
effects of evaporative cooling.

As we will see, the molecular velocity distributions obtained during the
cooling process are far from thermal. There are some molecules that never have
a collision during the whole simulation and so remain at high energy
throughout. Almost all these molecules have a kinetic energy greater than
10\,mK, and they disproportionately skew the mean kinetic energy of the sample
as a whole. Our interest is in the molecules that cool, and so we separate the
kinetic energy distribution into two parts, above and below 10\,mK. To express
how well the cooling works, we give the fraction of molecules in the low-energy
part, and their mean kinetic energy, both as functions of time.

\section{Collision models}
\label{Sec:CollisionModels}

In previous modeling \cite{Tokunaga:2011}, atoms and molecules collided like
hard spheres. In this model, the momenta in the center-of-mass frame before and
after a collision, ${\bf p}_{\text{c}}$ and ${\bf p}_{\text{c}}'$, are related
by
\begin{equation} {\bf p}_{\text{c}}' = {\bf p}_{\text{c}} - 2({\bf
p}_{\text{c}} \cdot {\bf \hat{e}}){\bf \hat{e}},
\end{equation}
where ${\bf \hat{e}}$ is a unit vector along the line joining the centers of
the spheres, given by
\begin{equation}
{\bf \hat{e}} ={\bf \hat{p}}_{\text{c}} \sqrt{1-|{\bf b}|^{2}} + {\bf b},
\end{equation}
where ${\bf \hat{p}}_{\text{c}}$ is a unit vector in the direction of ${\bf
p}_{\text{c}}$ and ${\bf b}$ is a vector that lies in a plane perpendicular to
${\bf p}_{\text{c}}$ and whose magnitude is the impact parameter divided by the
sum of the radii of the two spheres. For each collision, ${\bf b}$ is chosen at
random from a uniform distribution, subject to the constraints ${\bf b}\cdot
{\bf p}_{\text{c}} = 0$ and $|{\bf b}| \le 1$.

\begin{figure}[tbh!]
\centering
\includegraphics[width=0.45\textwidth]{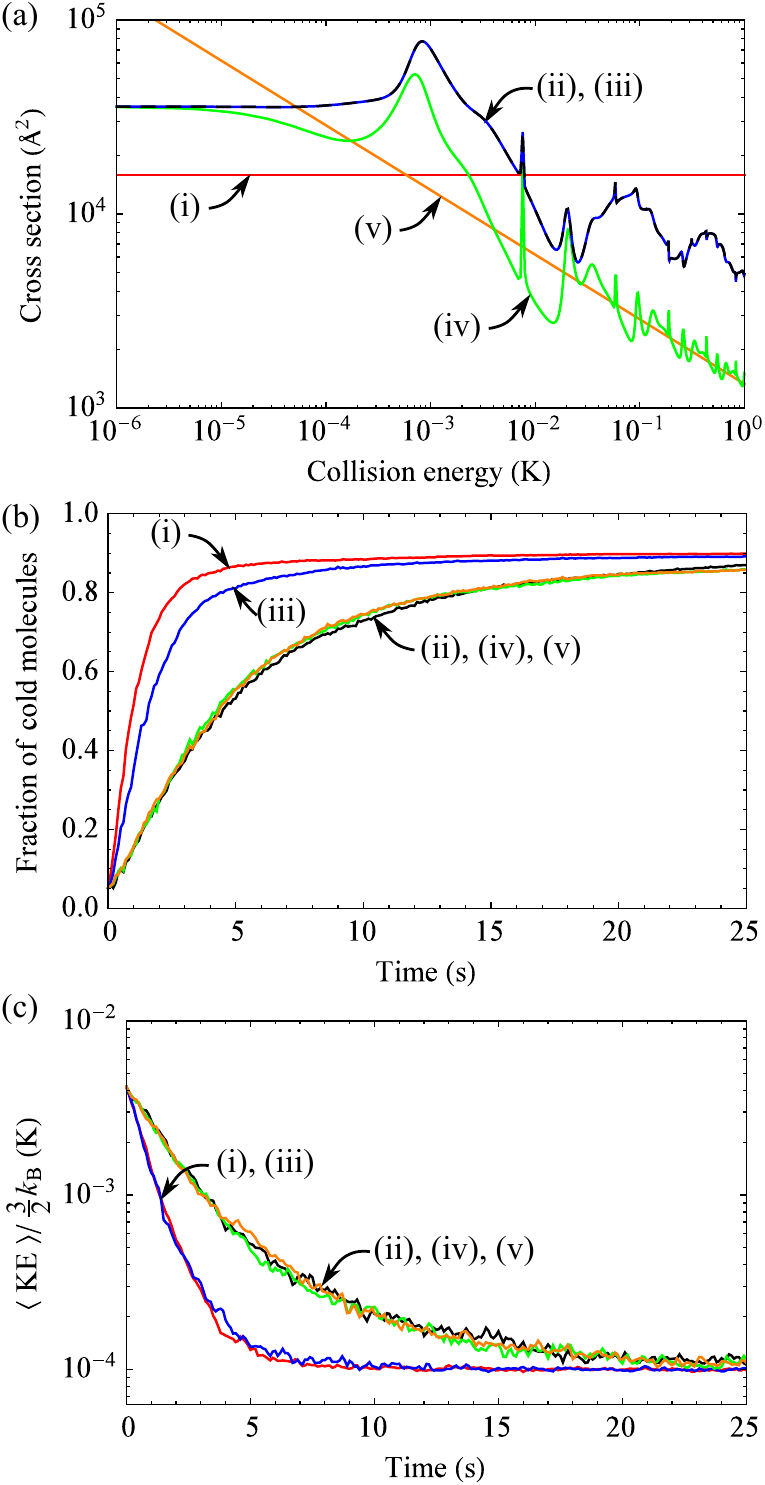}

\caption{\label{collisionmodel} (Color online) Results of various collision
models: (i) hard-sphere model with energy-independent cross section
$4\pi\bar{a}^2$; (ii) full energy-dependent differential cross section model;
(iii) hard-sphere model with $\sigma_{\text{el}}(E^\text{CM})$; (iv)
hard-sphere model with $\sigma_{\eta}^{(1)}(E^\text{CM})$; (v) hard-sphere
model with classical approximation to $\sigma_{\eta}^{(1)}(E^\text{CM})$. The
graphs show: (a) Cross section versus collision energy; (b) fraction of
molecules with kinetic energy below 10\,mK versus time; (c) mean kinetic energy
of that fraction versus time. The coolant is Rb and $a=+1.5\bar{a}$.}
\end{figure}

The lines labeled (i) in Fig.~\ref{collisionmodel} show how the cooling
proceeds for CaF + Rb when we use the hard-sphere model and choose the cross
section to be independent of energy and equal to $4\pi\bar{a}^2=1.59 \times
10^{-16}$\,m$^{2}$. The cross section is shown in Fig.~\ref{collisionmodel}(a),
while the cold fraction and the mean kinetic energy of that fraction are shown
in parts (b) and (c), both as functions of time. As explained in
Sec.\,\ref{sec:method}, the cold fraction is defined as the fraction with
kinetic energy below 10\,mK. The cold fraction increases rapidly, and that
fraction thermalizes quickly with the atoms. After just 4\,s, 85\% of the
molecules are in the cold fraction and their mean energy is within 50\% of the
100\,$\mu$K temperature of the coolant atoms.

The energy-independent hard-sphere (EIHS) model described above is reasonable
at very low energy, but it has three deficiencies. First, it neglects the fact
that the low-energy cross sections are actually $4\pi a^2$, where
$a$ is the true scattering length as opposed to the mean scattering
length. The true scattering length can take any value between $-\infty$ and
$+\infty$, but is generally unknown for a specific system until detailed
measurements are available to determine it. Secondly, the EIHS model neglects
the fact that real cross sections are strongly energy-dependent, usually
showing resonance structure on a background that drops off sharply with
increasing energy, as shown in Fig.\ \ref{collisionmodel}(a). Thirdly,
collisions with small deflection angles (forwards scattering) do not contribute
efficiently to cooling, and the EIHS model neglects the fact that differential
cross sections (DCS) at higher energies tend to be dominated by such forwards
scattering, because many collisions encounter only the attractive long-range
tail of the interaction potential.

To remedy all these deficiencies, we introduce here a new model that we call
the full DCS model. For this we calculate realistic integral and differential
cross sections, as described above, for a variety of choices of the scattering
length $a$. We use the elastic cross section $\sigma_{\text{el}}(E^\text{CM})$
from these calculations to determine the collision probability. This cross
section is curve (ii) in Fig.~\ref{collisionmodel}(a), and it is smaller than
in the EIHS model at collision energies above 8\,mK, but larger below 8\,mK. We
then select a deflection angle $\Theta$ from a random distribution that
reproduces the full differential cross section, $d\sigma/d\omega$, at energy
$E^\text{CM}$. To select a deflection angle at random from this distribution,
we form the cumulative distribution function,
\begin{equation}
S(\Theta) = \frac{2\pi}{\sigma_{\text{el}}}\int_{0}^{\Theta}
\frac{d\sigma}{d\omega} \sin(\Theta') d\Theta',
\end{equation}
select a random number $r$ between 0 and 1, and find the value of $\Theta$
where $S(\Theta) = r$.

The full DCS model is our most complete one and we have used it for all the
simulations in the following sections. Its results for the choice
$a=+1.5\bar{a}$ are shown by the lines labeled (ii) in
Fig.~\ref{collisionmodel}. It may be seen that the cooling proceeds more slowly
than in the EIHS model. It takes 14\,s for the cold fraction to reach 80\% and
for the energy of that fraction to be within 50\% of the temperature of the
atoms. The slower cooling is mainly due to the dominance of forward scattering
at higher energies.

There are three approximations to the full-DCS model that are worth considering
because they avoid the tabulation of differential cross sections and cumulative
distributions. The first of these is to use a hard-sphere collision model but
to take the full energy-dependent elastic cross section from Fig.\
\ref{collisionmodel}(a). This produces the cooling behavior labeled (iii) in
Figs.\ \ref{collisionmodel}(b) and (c). It may be seen that this
model produces cooling slightly slower than the EIHS model, but considerably
faster than the full DCS model. The second and more satisfactory approximation
is to use a hard-sphere collision model but to take the full energy-dependent
transport cross section $\sigma_\eta^{(1)}$, shown as line (iv) in Fig.\
\ref{collisionmodel}(a). We label this approach EDT-HS. It produces the
cooling behavior labeled (iv) in Figs.\ \ref{collisionmodel}(b) and (c). It may
be seen that it models the cooling of the molecules very accurately, because it
takes proper account of the reduced efficiency of small-angle collisions for
sympathetic cooling. However, as will be seen in Sec.\,\ref{Sec:HeatingAndLoss}, the EDT-HS approach does
{\em not} adequately model heating and loss of the coolant atoms.

It is worth exploring whether a classical calculation of $\sigma_{\eta}^{(1)}$
would suffice. Unlike the elastic cross section, $\sigma_{\eta{\rm
,class}}^{(1)}$ is finite because the factor of $1-\cos\Theta$ suppresses the
divergence due to forwards scattering. We have calculated $\sigma_{\eta{\rm
,class}}^{(1)}$ for the Lennard-Jones potentials described above,
\begin{equation}
\sigma_{\eta{\rm ,class}}^{(1)}=2\pi\int_0^\infty b[1-\cos\Theta(b)]db,
\end{equation}
where $b$ is the impact parameter and $\Theta(b)$ is the classical deflection
function \cite{Child:1974}. We find that it is very well approximated by the
power law $\sigma_{\eta}^{(1)}(E^\text{CM}) = A (E^\text{CM}/C_{6})^{-1/3}$,
with the dimensionless constant $A=4.79$. This cross section is labeled (v) in
Fig.~\ref{collisionmodel}(a). It agrees well with the quantum-mechanical
$\sigma_{\eta}^{(1)}(E^\text{CM})$ for Rb+CaF at high energies, as we would
expect when many partial waves contribute. Remarkably, the temperature and cold
fraction shown for this model in Fig.\ \ref{collisionmodel} agree very well
with those for model (ii), even as the temperature approaches 100\,$\mu$K. This
is an atypical result because, for $a=+1.5\bar{a}$, $\sigma_{\eta{\rm
,class}}^{(1)}$ is within a factor of about three of the quantum-mechanical
$\sigma_{\eta}^{(1)}$ at all energies above 3\,$\mu$K. For other values of $a$,
the two cross sections can differ by more than a factor of three at energies
below about $2\bar{E}$, which is around 1\,mK for Rb+CaF. Note that the
classical approximation will be less successful for a lighter coolant such as
Li where $\bar{E}$ is far higher.

\section{Approximate cooling rates}

\begin{figure}[tb]
\centering
\includegraphics[width=0.45\textwidth]{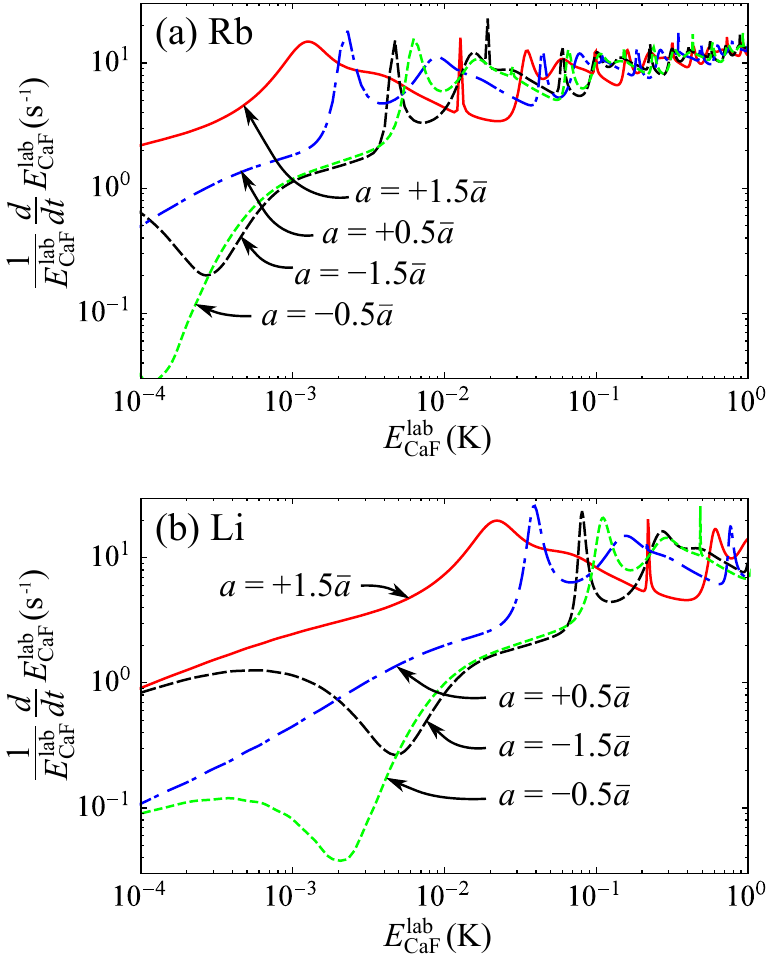}
\caption{\label{coolingrate} (Color online) Cooling rate of molecules as a
function of their kinetic energy, estimated from Eq.\,(\ref{dEdtApprox}), when
the coolant is (a) Rb and (b) Li, and for various values of the s-wave
scattering length: $a=+1.5\bar{a}$ (red solid line),
$a=+0.5\bar{a}$ (blue dash-dot line), $a=-0.5\bar{a}$
(green dotted line), and $a=-1.5\bar{a}$ (black dashed line).}
\end{figure}

From the transport cross sections, $\sigma^{(1)}_{\eta}$, in Fig.~\ref{crosssection} we
can make a useful estimate of the cooling rate of molecules as a function of
their kinetic energy. For this estimate, we assume stationary atoms with a
uniform density $n_{\text{at}}=10^{11}$\,cm$^{-3}$. The cooling rate is
\begin{equation}
\label{dEdtApprox}
\frac{d{E^\text{lab}_\text{CaF}}}{d t}={n_{\text{at}}}{\sigma}(E^\text{CM}) v \overline{{\Delta}E},
\end{equation}
where $v=(2E^\text{lab}_\text{CaF}/m_{\text{CaF}})^{1/2}$ is the
speed of the molecule and $\overline{{\Delta}E}$ is the average energy transfer
for a hard-sphere collision. $\Delta E$ is given explicitly as
\begin{equation}
\label{DeltaE}
\Delta E=-\left(\frac{2\mu}{m_{\rm CaF}+m_{\rm at}}\right)(1-\cos\Theta)E^\text{lab}_\text{CaF}.
\end{equation}

Figure~\ref{coolingrate} shows the cooling rates obtained this way, which
although only approximate are helpful for understanding the numerical results
presented later. For collisions with Rb at energies above 10\,mK, the cooling
rate does not depend strongly on the s-wave scattering length. This is the
energy regime where the $a$-independent classical approximation to
$\sigma_{\eta}^{(1)}(E^\text{CM})$ described in Sec.~\ref{Sec:CollisionModels}
is accurate. Due to the small reduced mass in the lithium case, the classical
limit is only reached for temperatures above 200\,mK, and so the cooling rate
depends sensitively on $a$ over the whole energy range of interest. When $a$ is
negative there is a minimum in the cooling rates corresponding to the
Ramsauer-Townsend minimum in $\sigma_{\eta}^{(1)}(E^\text{CM})$. For Rb, at
$a=-1.5\bar{a}$, this minimum is near 100\,$\mu$K, which is close to the
temperature of the atoms in our simulations and so will not have a significant
impact on the thermalization. For Li, the minimum occurs for kinetic energies
between 1 and 10\,mK, and so it has a strong effect on the thermalization.
Finally, we note that in the ultracold limit the cooling rate is almost an
order of magnitude higher for Rb than for Li, reflecting the larger value of
$\bar{a}$ for Rb + CaF relative to Li + CaF.

\section{Cooling dynamics}
\label{Sec:CoolingDynamics}

\begin{figure}[tb]
\centering
\includegraphics[width=0.48\textwidth]{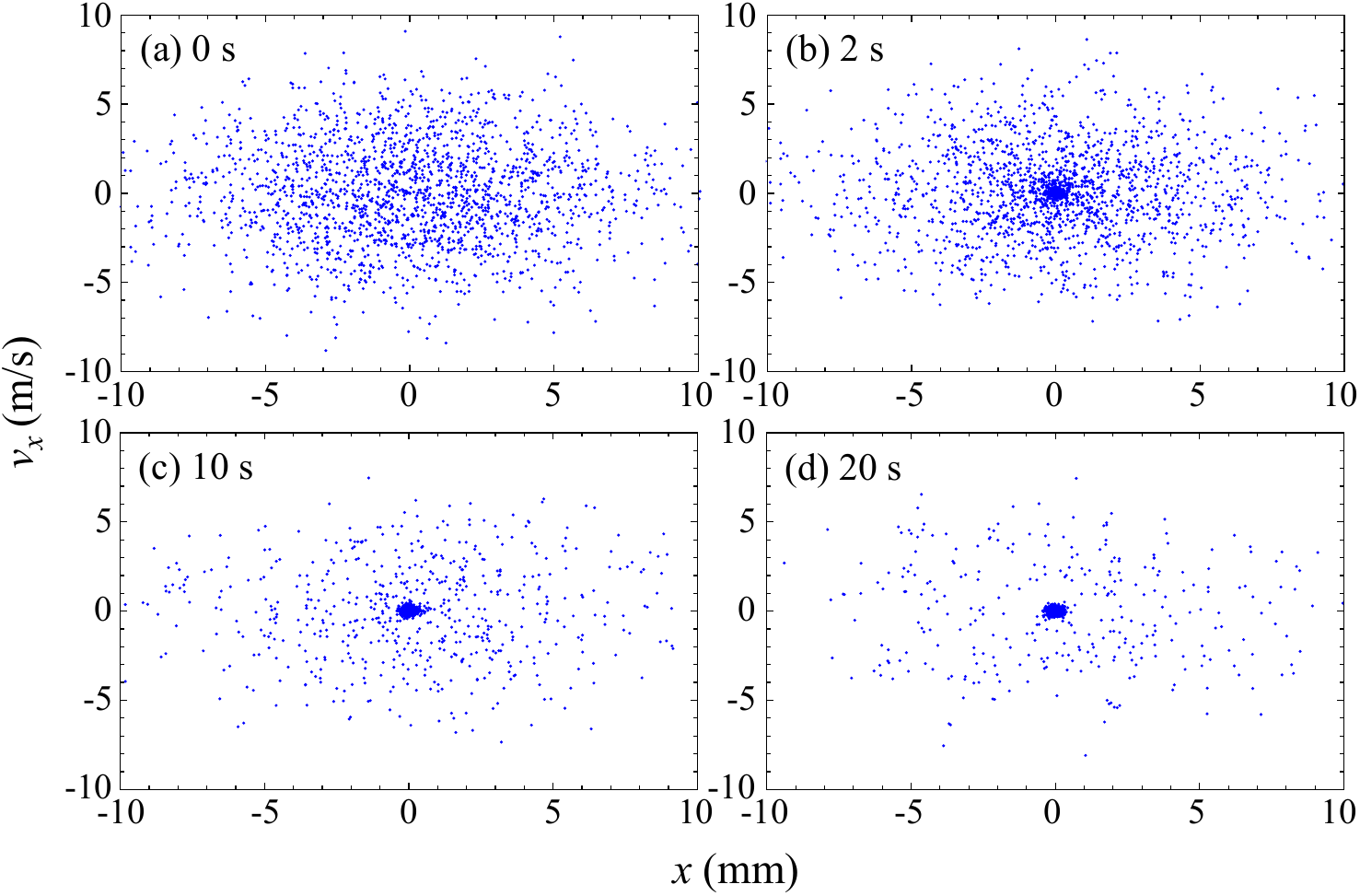}
\caption{\label{phasespace}(Color online) Time evolution of the phase-space
distribution of molecules in the $x$ direction. The cooling times are (a) 0\,s,
(b) 2\,s, (c) 10\,s, and (d) 20\,s. The coolant is Rb and
$a=+1.5\bar{a}$.}
\end{figure}

Figure~\ref{phasespace} shows the evolution of the $(x,v_{x})$ phase-space
distribution of CaF when Rb atoms are used as the coolant, for the case where
the s-wave scattering length is $a=+1.5\bar{a}$. At $t=0$
(Fig.~\ref{phasespace}(a)), the molecules fill the phase-space acceptance of
the trap. At later times, more and more molecules congregate at the trap center
as they are cooled by collisions with the atoms. After 20\,s
(Fig.~\ref{phasespace}(d)), the distribution has separated into two parts. The
majority are cooled to the center, but there are some that remain uncooled.
These are molecules that have large angular momentum around the trap center and
so are unable to reach the center where the atomic density is high. At
$x=3$\,mm for example, the atomic density, and hence the collision rate, is a
factor of 1000 smaller than at the center, and so molecules at this distance
are unlikely to collide with atoms on the 20\,s timescale shown in the figure.
These molecules can be cooled by expanding the size of the atom cloud, but only
at the expense of the overall cooling rate \cite{Tokunaga:2011}.

\begin{figure}[tb]
 \centering
 \includegraphics[width=0.48\textwidth]{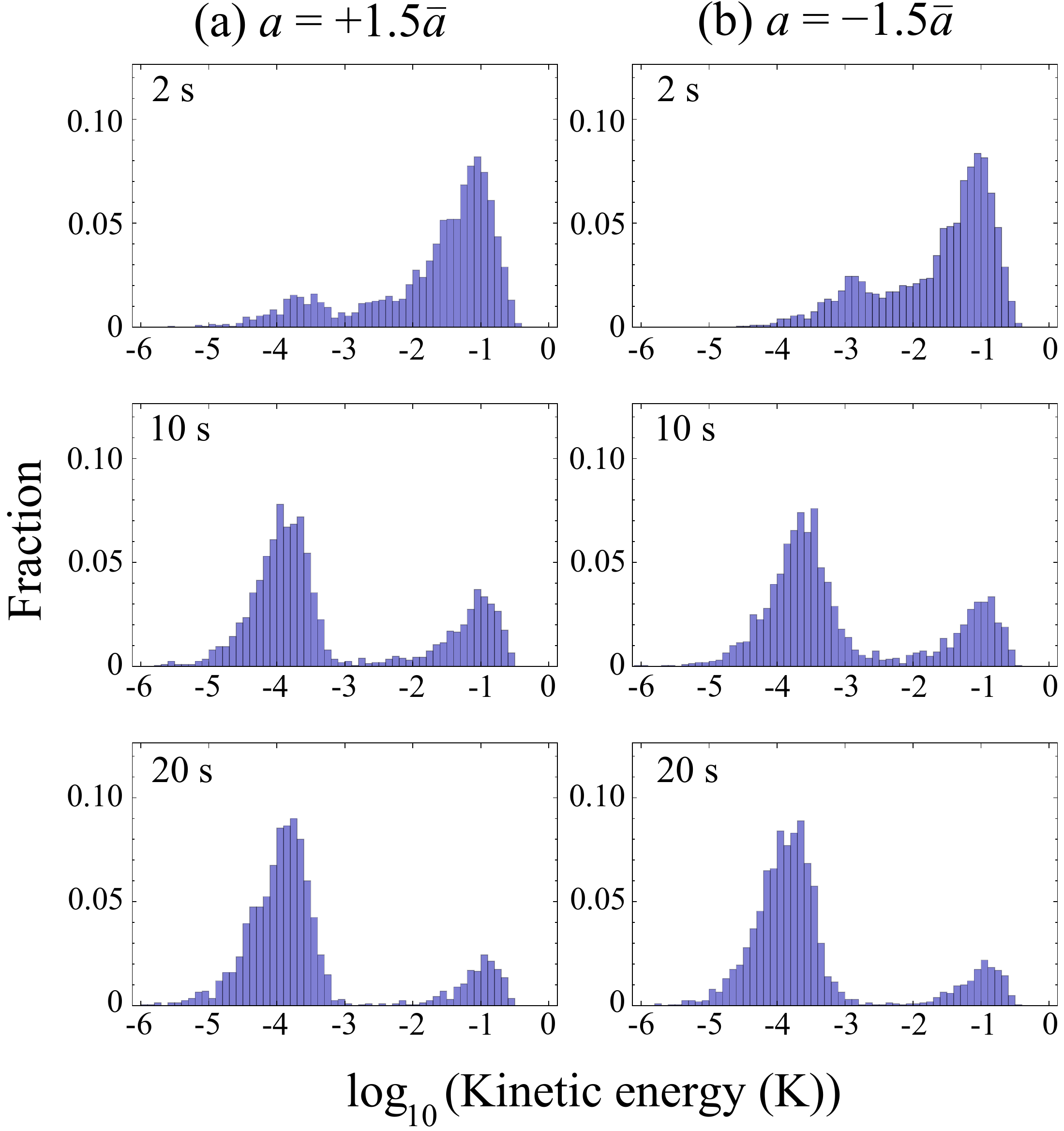}
\caption{\label{rbhistogram}(Color online) Kinetic energy distributions after
2\,s, 10\,s, and 20\,s. The coolant is Rb. Left panels have
$a=+1.5\bar{a}$ while right panels have $a=-1.5\bar{a}$.}
\end{figure}

Figure \ref{rbhistogram}(a) shows histograms of the kinetic energy distribution
of the molecules at three different times, 2, 10 and 20\,s, when the coolant is
Rb and $a=+1.5\bar{a}$. These are the same times as chosen for the phase space
distributions in Fig.\,\ref{phasespace}, and the results come from the same
simulation. The initial distribution is a Maxwell-Boltzmann distribution with a
temperature of 70\,mK, truncated at the trap depth of 400\,mK. The distribution
rapidly separates into two parts, those that cool and those that do not. The
latter are the molecules that never reach the trap center because of their
large angular momentum, as discussed above. A significant fraction of molecules
are cooled below 1\,mK after just 2\,s. After 10\,s the majority are in this
group, and after 20\,s this cold fraction is almost fully thermalized with the
atoms. We return to part (b) of Fig.\,\ref{rbhistogram} in the next section.
\section{Sensitivity to the scattering length and the choice of coolant}
\label{sec:sensitivity}

At low energies, cross sections are very sensitive to the exact form of the
scattering potential, as shown in Fig.\ \ref{crosssection}, and cannot be
calculated accurately without independent knowledge of the scattering length.
In our model Lennard-Jones potential, the full energy-dependence of the cross
section is determined once the s-wave scattering length, $a$, is
fixed. Here, we study how the simulation results change as we vary the value of
$a$. The choice of coolant is also a crucial consideration, and so
we compare the results for Li and Rb as coolants.

\subsection{Evolution of the kinetic energy distributions}

Figure \ref{rbhistogram} compares how the kinetic energy distributions evolve
for two cases: $a=+1.5\bar{a}$ and $a=-1.5\bar{a}$, with
Rb as the coolant. At 2\,s the two distributions are similar. The main
difference is that the distribution extends to lower energies for
$a=+1.5\bar{a}$. The similarity is due to the similar cooling rates
at the high energies, as shown in Fig.\,\ref{coolingrate}(a), while the
difference at low energy is due to the far higher cooling rate for
$a=+1.5\bar{a}$ at energies below 1\,mK (compare the solid red and
black dashed lines in Fig.\,\ref{coolingrate}(a)). Exactly the same trend is
seen after 10\,s of cooling. Once again, the high-energy parts of the
distributions are very similar, but the distribution extends to lower energies
for the $a=+1.5\bar{a}$ case. After 20\,s the majority of the
molecules have fully thermalized with the atoms and the two distributions are
very similar to one another.

\begin{figure}[tb]
 \centering
 \includegraphics[width=0.48\textwidth]{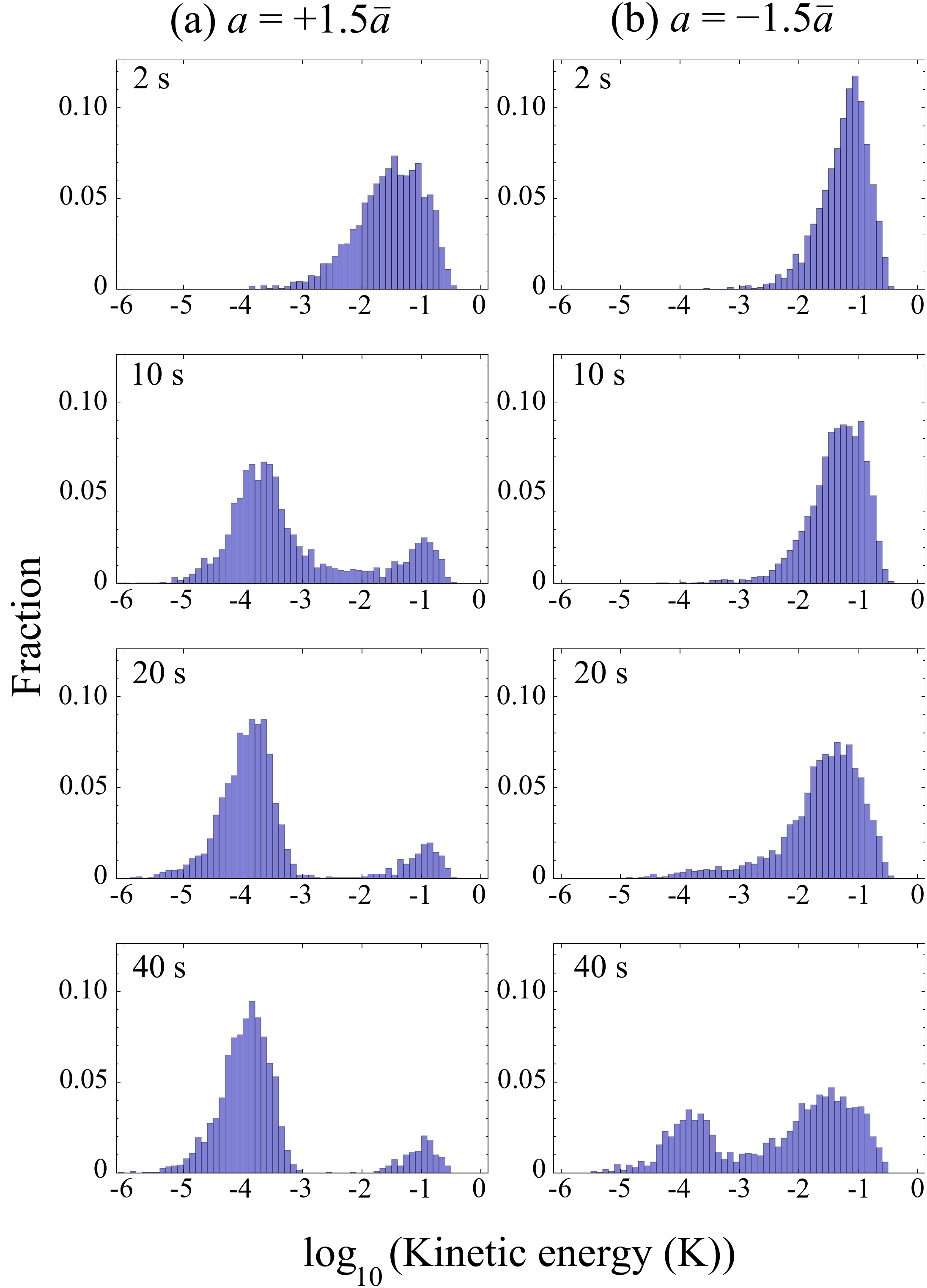}
\caption{\label{lihistogram}(Color online) Kinetic energy distributions after
2\,s, 10\,s, 20\,s and 40\,s. The coolant is Li. Left panels have
$a=+1.5\bar{a}$ while right panels have $a=-1.5\bar{a}$.}
\end{figure}

Figure \ref{lihistogram} shows the corresponding histograms for the case of Li.
Here, the cooling proceeds more slowly and so we have added a fourth pair of
histograms showing the distributions after 40\,s. There is a great contrast
between the positive and negative scattering lengths in this case. For
$a=+1.5\bar{a}$ the distribution evolves in a very similar manner to
the Rb case, but when $a=-1.5\bar{a}$ it takes a long time for the
molecules to reach energies below 10\,mK. This is the effect of the
Ramsauer-Townsend minimum which reduces the cooling rate estimated in
Fig.\,\ref{coolingrate}(b) to 0.25\,s$^{-1}$ for kinetic energies near 20\,mK.
Because the minimum is broad in energy, and there is a large mass mismatch
between CaF and Li, a collision cannot take a molecule directly across the
minimum. The molecules have to be cooled \textit{through} the minimum by
multiple collisions, and that takes a long time. Once molecules have passed
through this minimum, cooling to ultracold temperatures occurs on a similar
timescale to the $a=+1.5\bar{a}$ case.

\subsection{Cold fraction and mean kinetic energy}

\begin{figure}[tb]
 \centering
 \includegraphics[width=0.45\textwidth]{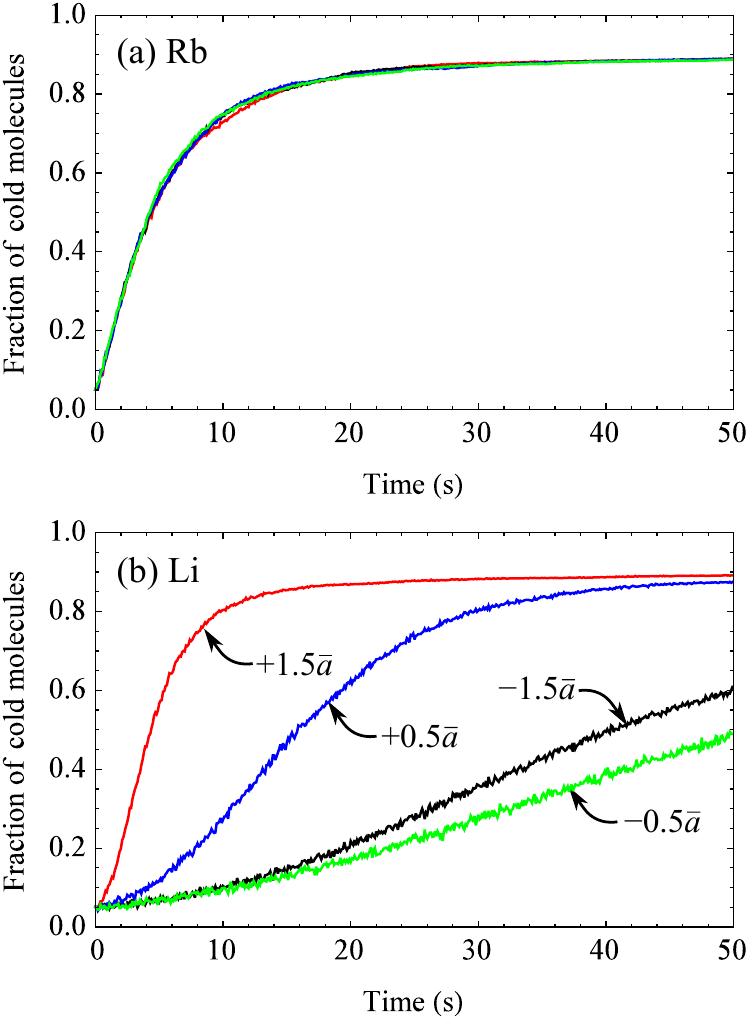}
\caption{\label{fraction}(Color online) Fraction of molecules with kinetic
energy below 10\,mK as a function of time for (a) Rb, and (b) Li, for four
different values of the scattering lengths: $a=+1.5\bar{a}$ (red),
$a=+0.5\bar{a}$ (blue), $a=-0.5\bar{a}$ (green), and $a=-1.5\bar{a}$ (black).}
\end{figure}

Figure~\ref{fraction}(a) shows the fraction of molecules with kinetic energy
less than 10\,mK, as a function of time, for various values of $a$ when the
coolant is Rb. This fraction is entirely insensitive to $a$. This is because
the cooling rate is independent of $a$ for energies above 10\,mK, as we saw in
Fig.\,\ref{coolingrate}. After 5\,s about 50\% of the molecules are in this
cold fraction, and after 20\,s this exceeds 80\%. Figure~\ref{fraction}(b)
shows the cold fraction versus time when the coolant is Li. We find a strong
dependence on $a$ in this case. When $a=+1.5\bar{a}$, the increase in the cold
fraction with time is similar to the Rb case. For this value of $a$ there is a
maximum in the cooling rate at a kinetic energy of about 70\,mK (see
Fig.\,\ref{coolingrate}(b)), which happens to match the initial temperature of
the molecules, and so the cooling to below 10\,mK proceeds rapidly. The cold
fraction reaches 50\% after 4\,s in this case. The increase in the cold
fraction is slower for $a=+0.5\bar{a}$, reaching 50\% after 16\,s. The
accumulation of cold molecules is exceedingly slow when $a$ is negative. When
$a=-1.5\bar{a}$, the Ramsauer-Townsend minimum is at $E^\text{lab}_\text{CaF}=
20$\,mK, and it takes a long time for the molecules to cool through this
minimum. The cold fraction reaches 50\% after 40\,s in this case. When
$a=-0.5\bar{a}$, the Ramsauer-Townsend minimum is shifted to
$E^\text{lab}_\text{CaF}= 10$\,mK, but the cross section at the minimum is a
factor of five smaller, and so the cooling is even slower, taking 50\,s to
reach 50\%.

\begin{figure}[tb]
 \centering
 \includegraphics[width=0.45\textwidth]{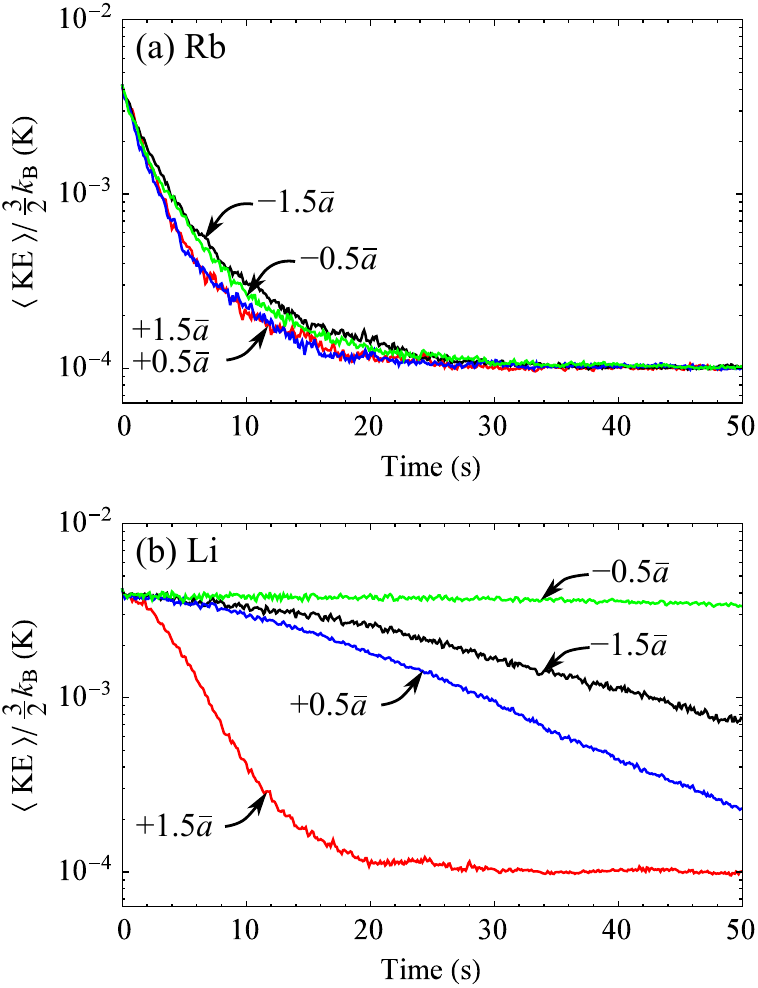}
\caption{\label{temperature} (Color online) Mean kinetic energy of the cold
fraction as a function of time when the coolant is (a) Rb and (b) Li, and for
various values of the s-wave scattering length: $a=+1.5\bar{a}$
(red), $a=+0.5\bar{a}$ (blue), $a=-0.5\bar{a}$ (green),
and $a=-1.5 \bar{a}$ (black).}
\end{figure}

Figure \ref{temperature}(a) shows the mean kinetic energy of the cold fraction
as a function of time for various values of $a$ when Rb is used as
the coolant. As for the cold fraction itself, this measure is almost
independent of $a$. This may seem surprising, since the cooling
rates estimated in Fig.\,\ref{coolingrate}(a) show a strong dependence on
$a$ below a few mK. However, the mean kinetic energy is strongly
influenced by molecules with kinetic energies close to the 10\,mK cutoff that
defines the cold fraction, and at this energy the cooling rates show little
dependence on $a$. We find a small difference in the cooling rates
between positive and negative scattering lengths. For the positive
$a$ values the molecular temperature is within a factor of two of
the atomic temperature after 10\,s, while for the negative $a$
values this takes 14\,s. Figure \ref{temperature}(b) shows how the mean
kinetic energy of the cold fraction evolves when Li is used as a coolant. In
this case, the cooling depends sensitively on $a$. When
$a=+1.5\bar{a}$ the evolution is similar to the Rb case. The cooling
is much slower when $a=+0.5\bar{a}$ because the low-energy
cross-section is nine times smaller. The cooling is even slower when
$a$ is negative. This is because, in the energy region between 1 and
10\,mK, the Ramsauer-Townsend minimum greatly suppresses the cooling rate
relative to the positive $a$ case, and because molecules with
energies in this range have a strong influence on the mean.

\begin{figure}[tb]
 \centering
 \includegraphics[width=0.45\textwidth]{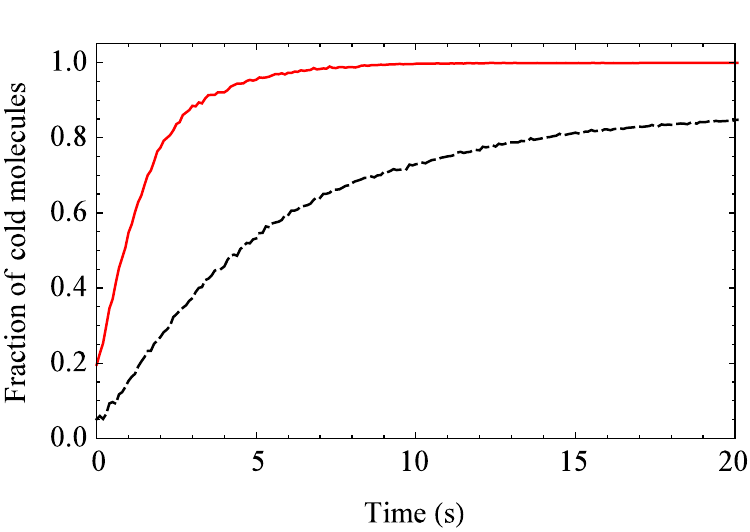}
\caption{\label{fraction20mK}(Color online) Fraction of cold molecules as a
function of time for the initial temperatures of $T_{\text{i}}=20$\,mK (red
solid line), and $T_{\text{i}}=70$\,mK (black dashed line). The coolant is Rb
and $a=+1.5\bar{a}$.}
\end{figure}

The fraction of molecules that are
cooled below 10\,mK depends on the initial temperature, $T_{\text{i}}$.
Figure~\ref{fraction20mK} compares this fraction for $T_{\text{i}}=20$\,mK and
70\,mK, for the case where Rb is the coolant and $a=+1.5\bar{a}$.
These two initial temperatures correspond to temperatures of 2\,mK and 30\,mK
prior to compression of the cloud in the microwave trap. When
$T_{\text{i}}=20$\,mK, more than 99\% of the molecules are cold within 10\,s.

\section{Atom heating and loss}
\label{Sec:HeatingAndLoss}

\begin{figure}[tb]
 \centering
 \includegraphics[width=0.47\textwidth]{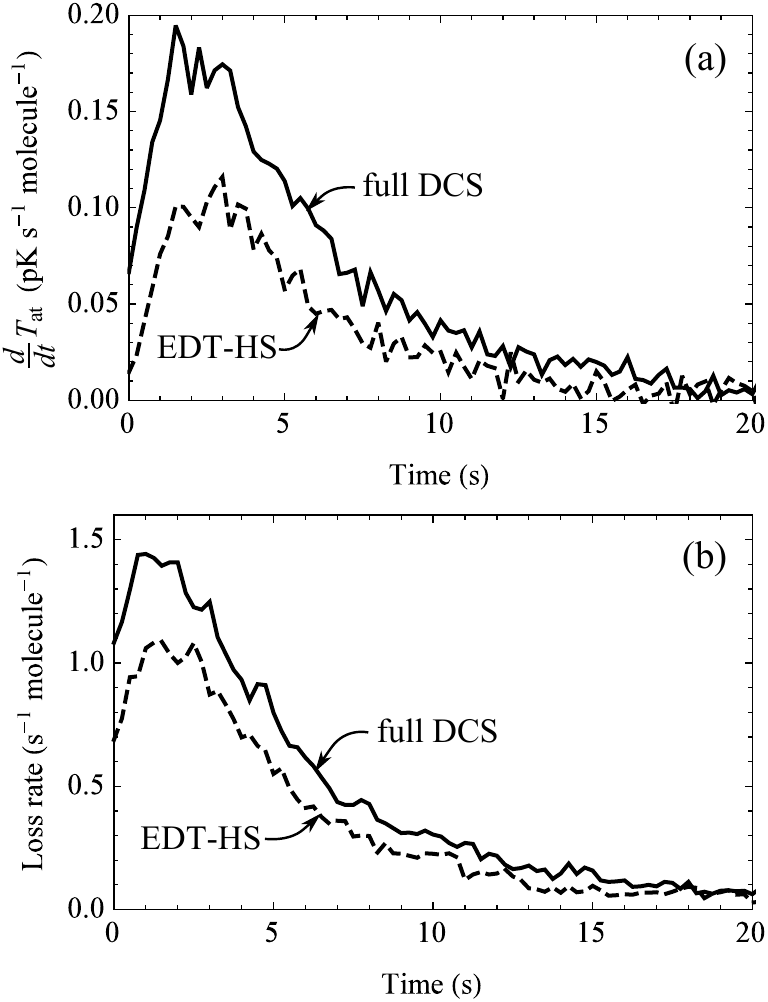}
\caption{\label{lossandheating} (a) Atom heating rate per molecule and (b) atom
loss rate per molecule for the EDT-HS model (dashed line) and the full DCS
model (solid line). The coolant is Rb, $a=+1.5\bar{a}$, and the molecules have
an initial temperature of 70\,mK. There are $10^{5}$ molecules and $10^{9}$
atoms.}
\end{figure}

The energy transferred from molecules to atoms will either eject atoms from the
trap, or will heat them up. As described in Sec.\,\ref{sec:method}, we suppose
that atoms are lost from the trap if their total energy exceeds 1\,mK. This
could be the actual depth of the trap, or an ``rf knife'' might be used to cut
off the trap at this depth. Here, we investigate the heating and loss of atoms
and the consequences for sympathetic cooling. We note that while the EDT-HS
collision model correctly captures the molecule cooling dynamics when
$\sigma_{\eta}^{(1)}$ is used as the cross section, it does not model correctly
the atom heating and loss. Here, we highlight the difference between these two
approaches by comparing the results obtained from the EDT-HS model and the full
DCS model.

Figure~\ref{lossandheating} shows how the heating and loss rates of the atoms
change with time in the full DCS model and the EDT-HS model for the case of
$10^5$ molecules and $10^9$ atoms. The two models show similar trends, so we
first discuss these trends and then consider the differences between the
models. At early times the majority of the molecules have energies far above
the atom trap depth and so most collisions cause atom loss, rather than
heating. The loss rate is high while the heating rate is low. Nevertheless,
there is still some heating due to small-angle collisions with the molecules
which transfer only a little energy to the atoms. The loss rate increases
during the first second because the collision cross section and the
atom-molecule overlap both increase as the molecules are cooled. As time goes
on the loss rate falls because the molecules are cooler and there are fewer
collisions with enough energy to kick atoms out of the trap. For the same
reason the heating rate initially increases, but then decreases again as the
molecules have less energy to transfer to the atoms. For most of the 20\,s
period, the full DCS model gives more atom heating and more atom loss than the
EDT-HS model. Only at long times, once the atoms and molecules are almost fully
thermalized, do the two models give the same results.

Integrating the results of the full DCS model shown in
Fig.\,\ref{lossandheating}, we find that the total temperature increase of the
trapped atoms is 1.3\,pK per molecule, while the total loss is 10 atoms per
molecule. The energy deposited into the trapped atom cloud is only 1.8\% of the initial
energy of the molecular cloud. In this sense, the sympathetic cooling process
is remarkably efficient.

We now turn to how the atom heating and loss rates can be {\em understood}, and
explain why the two models give different results. Whether an atom is heated or
lost depends on the kinetic energy kick it receives in the collision, as given
by Eq.\,(\ref{DeltaE}) if the atoms are assumed to be stationary. An atom at
the center of the trap is lost from the trap if the energy transferred in the
collision exceeds the trap depth, $\Delta E > E_{\text{trap}}$. This occurs if
the deflection angle exceeds a critical angle $\Theta_{\text{crit}}$ given by
\begin{equation}
\cos\Theta_{\text{crit}} = 1 - \left(\frac{m_{\rm CaF}+m_{\rm at}}{2\mu}\right)
\left(\frac{E_{\text{trap}}}{E^\text{lab}_\text{CaF}}\right).
\end{equation}
At laboratory-frame energies below critical energy $E_{\rm crit} = (m_{\rm
CaF}+m_{\rm at})/4\mu)E_{\text{trap}}$, no loss is possible, assuming
stationary atoms at the center of the trap. This energy is 2.63 mK for Li+CaF
and 1.04 mK for Rb+CaF. All collisions below this energy and collisions above
this energy where $\Theta < \Theta_{\text{crit}}$ will not eject atoms from the
trap, but still transfer energy and so heat the atom cloud, by an amount
proportional to $1-\cos\Theta$. This suggests the possibility of defining cross
sections for atom heating and loss as partial integrals of the differential
cross section,
\begin{equation}
\label{sigmaloss}
\sigma_{\text{loss}}=2\pi\int_{-1}^{\cos\Theta_{\text{crit}}} \frac{d\sigma}{d\omega}
d\cos\Theta, \\
\end{equation}
\begin{equation}
\label{sigmaheat}
\sigma_{\text{heat}}=2\pi\int_{\cos\Theta_{\text{crit}}}^{1} \frac{d\sigma}{d\omega}
(1-\cos\Theta) d\cos\Theta.
\end{equation}
It is convenient to write these as integrals over $d\cos\Theta$ instead of
$\sin\Theta\,d\Theta$ because the $\cos\Theta$ form allows us to show plots in
which the integrals are simply areas that can be estimated by eye. Note that if
$E^\text{lab}_\text{CaF}<E_{\rm crit}$ then $\sigma_{\text{heat}} =
\sigma_{\eta}^{(1)}$, because all collisions cause heating rather than loss.
For the full DCS model these integrals must be evaluated numerically, but in
the hard-sphere model the DCS are isotropic, $d\sigma_{\rm HS}/d\omega =
\sigma_{\rm HS}/(4\pi)$, and the integrals can be evaluated analytically to
give
\begin{equation}
\sigma_{\rm loss,HS}=\frac{1}{2}(1+\cos\Theta_{\rm crit}) \sigma_{\rm HS}
\end{equation}
and
\begin{equation}
\sigma_{\rm heat,HS}=\frac{1}{4}(1-\cos\Theta_{\rm crit})^2 \sigma_{\rm HS}.
\end{equation}

\begin{figure}[htb]
 \centering
 \includegraphics[width=0.47\textwidth]{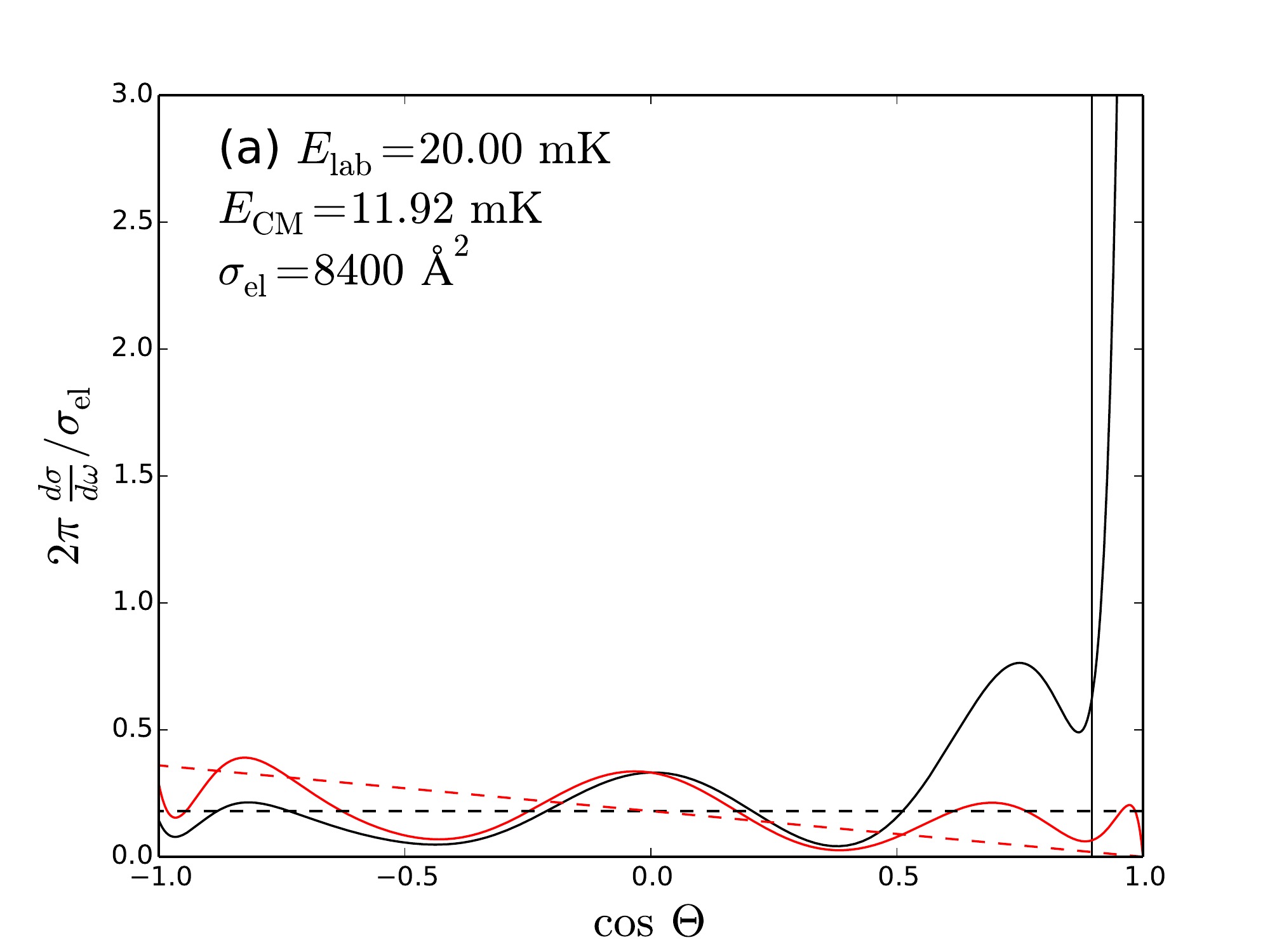}
 \includegraphics[width=0.47\textwidth]{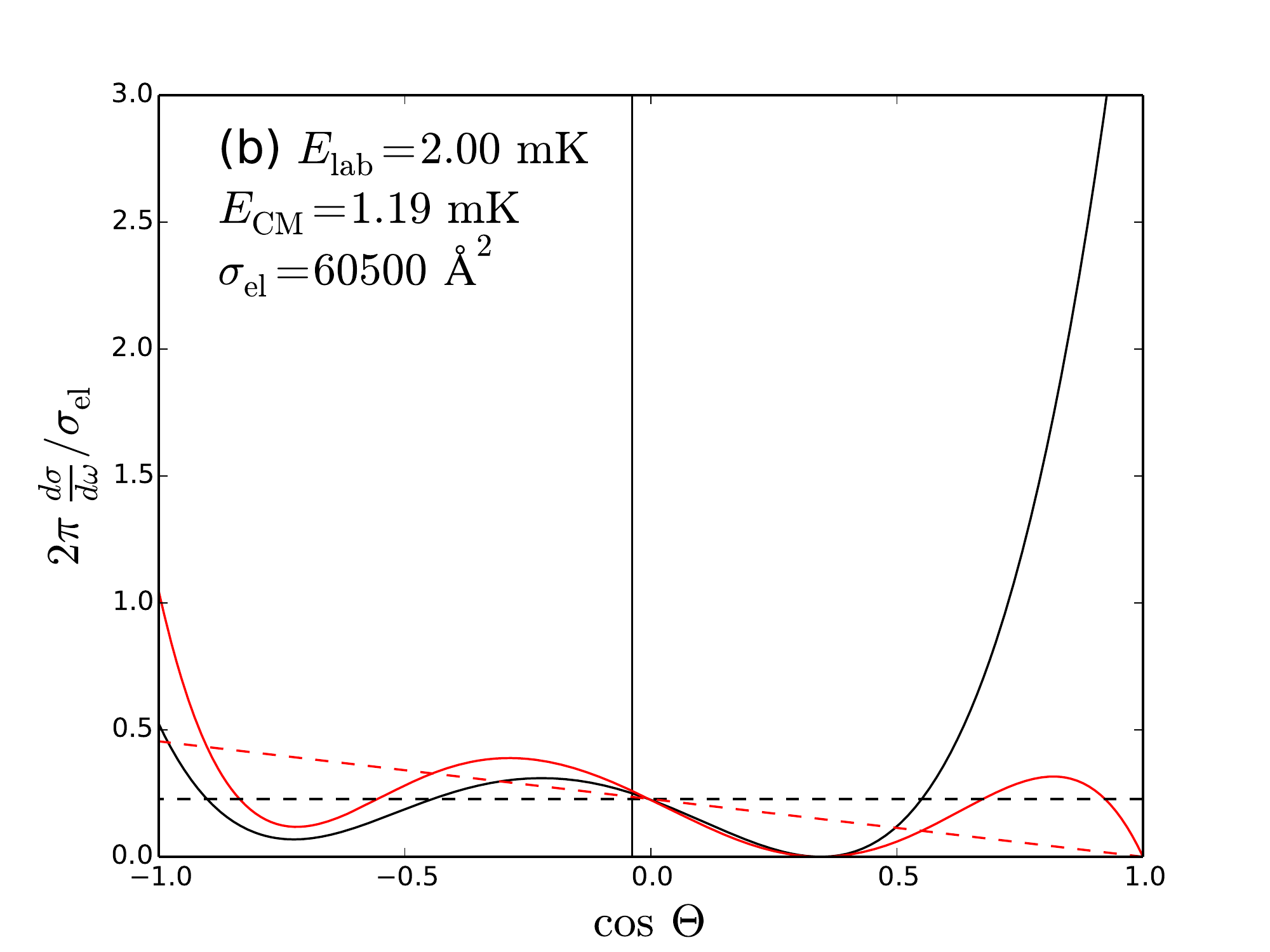}

\caption{\label{dcsplots} (Color
online) Differential cross sections and their contributions to heating and
loss. The solid black line shows the full quantum-mechanical $d\sigma/d\omega$,
while the solid red line shows $(1-\cos\Theta)d\sigma/d\omega$; the dashed
lines show the corresponding quantities for the EDT-HS model. The vertical line
shows the value of $\Theta_{\text{crit}}$. (a) $E^\text{lab}_\text{CaF} =
2$\,mK. (b) $E^\text{lab}_\text{CaF} = 20$\,mK. The coolant is Rb and
$a=+1.5\bar{a}$.}
\end{figure}

Figure \ref{dcsplots} shows differential cross sections at two energies that
correspond to $E^\text{lab}_\text{CaF} = 2$ mK and 20 mK for Rb+CaF. Both full
differential cross sections and those from the EDT-HS model are shown (solid
and dashed black lines respectively), and the corresponding quantities weighted
by $1-\cos\Theta$ are shown in red. The values of $\Theta_{\rm crit}$ at the
two energies are shown as vertical lines. Integrals over the complete range of
$\cos\Theta$ under the black lines correspond to $\sigma_{\rm el}$, and under
red lines correspond to $\sigma_\eta^{(1)}$; the latter is the same for the
full DCS and EDT-HS models by construction. $\sigma_{\rm loss}$ is the area
under the black lines to the left of $\Theta_{\rm crit}$, and $\sigma_{\rm
heat}$ is the area under the red lines to the right of $\Theta_{\rm crit}$. It
can be seen that at 20 mK the full DCS has a very large forwards peak; this
dominates $\sigma_{\rm heat}$, even though its contribution is suppressed by
the $1-\cos\Theta$ weighting. The resulting $\sigma_{\rm heat}$ is many times
larger than in the EDT-HS model, which has no forward peak. The full DCS also
has a secondary peak near $\cos\Theta=0.75$, which is outside $\Theta_{\rm
crit}$ and so contributes to atom loss; the resulting $\sigma_{\rm loss}$ is
also larger than in the EDT-HS model. At the lower energy of 2 mK, $\Theta_{\rm
crit}$ is near $\Theta=\pi/2$. There is still a large forwards peak but it no
longer dominates due to the changed range of integration, leading to similar
cross sections for the two models.

\begin{figure}[tb]
 \centering
 \includegraphics[width=0.47\textwidth]{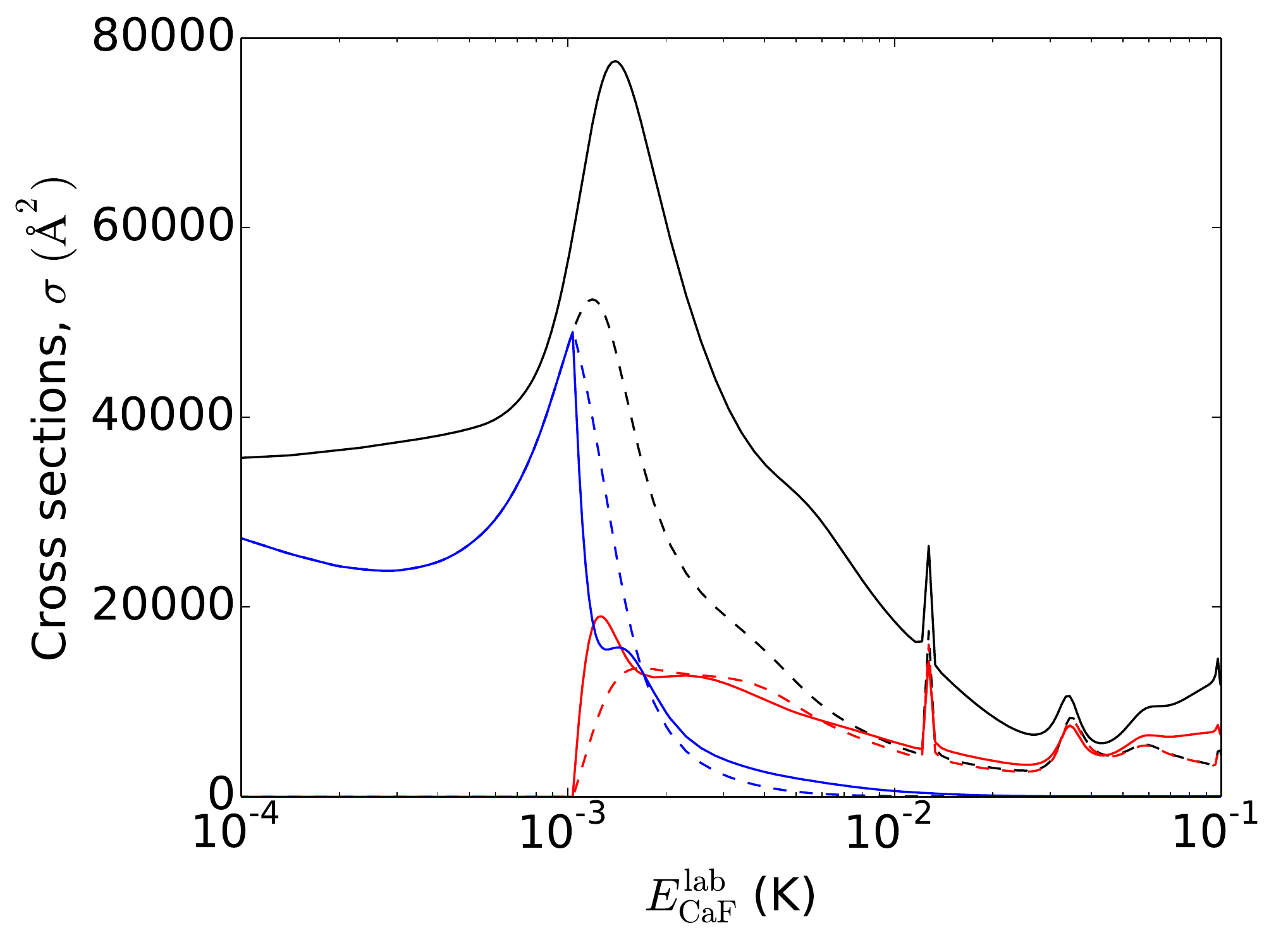}
\caption{\label{heat_loss_cross_sec} (Color online) Loss (red) and heating
(blue) cross sections as a function of CaF laboratory energy for the EDT-HS
model (dashed lines) and the full DCS model (solid lines). $\sigma_\text{el}$
(solid black line) and $\sigma_\eta^{(1)}$ (dashed black line) are shown for
comparison. The coolant is Rb and $a=+1.5\bar{a}$. }
\end{figure}

Figure \ref{heat_loss_cross_sec} shows how the heating and loss cross sections
vary over the range of energies relevant to the cooling process. As explained
above, at low energy, $E^\text{lab}_\text{CaF}<E_\text{crit}$, we have
$\sigma_\text{heat}=\sigma_\eta^{(1)}$ and $\sigma_\text{loss}=0$. Above
$E_\text{crit}$ the heating cross section falls off rapidly; for the EDT-HS
model it falls to negligibly small values by a few mK. The cross section for
the full DCS is several times larger than that for the EDT-HS model in this
tail, but it also falls towards zero. The loss cross sections for the two
models agree surprisingly well ($\pm\sim30\%$) in an intermediate energy range
from about 2\,mK to 60\,mK; the extent of this similarity is greatest for this
particular scattering length ($a=+1.5\bar{a}$), but it also exists up to about
20\,mK for the other scattering lengths investigated. Above this intermediate
range, $\sigma_\text{loss}$ for the full DCS model does become larger than for
the EDT-HS model, as we expect. The large peak around 1.5 mK in the elastic
cross section is a d-wave feature that causes a large amount of backwards
scattering around that energy; this significantly enhances the loss cross
section because at this energy $\Theta_\text{crit}$ is still near backwards
scattering.

The overall effect is that the full DCS model gives significantly larger rates
of both atom heating and atom loss than the EDT-HS model, especially at higher
energies, exactly as we see in Fig.\,\ref{lossandheating}. This is at first
sight surprising because each atom-molecule collision causes either atom
heating or atom loss. However, at higher energies the total collision rate is
considerably greater in the full DCS model than in the EDT-HS model, because
the former is determined by $\sigma_{\rm el}$ and the latter by
$\sigma_\eta^{(1)}$.

The effects of atom heating and loss will, of course, be most significant when
the atom number does not greatly exceed the molecule number. Table
\ref{moleculenumber} shows the results of simulations for a variety of molecule
numbers, with the atom number fixed at $10^{9}$, and once again compares the
full DCS and EDT-HS models. In the first three rows, the trap depth for the
atoms is 1\,mK. When the atom number is 100 times the molecule number, atom
heating and loss are not significant effects. For each molecule, the first few
collisions carry away most of the energy, and almost all of these collisions
cause atom loss, rather than heating. Thus, for this case, 11\% of the atoms
are lost, and the atom cloud heats up by just 13\,$\mu$K. The molecules
thermalize completely with the atoms, and the majority are in the cold
fraction. When the atom number is only 10 times the molecule number, the
effects are far more dramatic. At the end of the simulation only 2.2\% of the
atoms remain, and the temperature of those remaining has increased to
259\,$\mu$K. Since there are so few atoms remaining, only 70\% of the molecules
now reach kinetic energies below 10\,mK, and the temperature of this fraction
is increased to 596\,$\mu$K. The EDT-HS collision model underestimates the atom
loss and atom heating, and it predicts more cold molecules, with a lower final
temperature, than the full DCS model.

It is interesting to explore whether the atomic trap depth of
1\,mK used in the simulations above is optimum. The last three rows of Table
\ref{moleculenumber} show the results of simulations with the atomic trap depth
increased to 5\,mK. As expected, this results in less atom loss and more atom
heating. The fraction of cold molecules increases a little, but the temperature
of the cold fraction increases significantly. This is especially evident when
the atom number is only 10 times the molecule number. It is clear that large
atomic trap depths are not necessarily beneficial for sympathetic cooling, and
indeed there might be advantages in adjusting the trap depth as cooling
proceeds.

\begin{table}[tb]
\caption{\label{moleculenumber} The effect of different molecule numbers
($N_{\text{mol}}$), with atom number fixed at $10^{9}$, for two values of the
trap depth $E_{\text{trap}}$: 1\,mK and 5\,mK. The columns give the fraction of
remaining atoms $f_{\text{at}}$, the atomic temperature $T_{\text{at}}$, the
fraction of cold molecules $f_{\text{mol}}$, and the molecular temperature
$T_{\text{mol}}$ after 50\,s. The main values are for the full DCS model, and
the values in brackets are for the EDT-HS model.} \centering
\begin{tabular}{c|c|cccc}
\hline
\hline
$E_{\text{trap}}$ & $N_{\text{mol}}$ & $f_{\text{at}}$ (\%) & $T_{\text{at}}$ (${\mu}$K)
& $f_{\text{mol}}$ (\%) & $T_{\text{mol}}$ (${\mu}$K)\\
\hline
& $10^7$ & 89 (92) & 113 (107) & 89 (89) & 113 (108)\\
1\,mK & 5${\times}10^7$ & 38 (59) & 159 (136) & 88 (88) & 168 (144)\\
& $10^8$ & 2.2 (18) & 259 (180) & 70 (83) & 596 (246)\\
\hline
& $10^7$ & 95 (96) & 151 (133) & 90 (89) & 153 (134)\\
5\,mK & 5${\times}10^7$ & 75 (79) & 396 (291) & 90 (91) & 435 (299)\\
& $10^8$ & 50 (57) & 704 (518) & 85 (87) & 927 (624)\\
\hline
\hline
\end{tabular}
\end{table}

\section{The effect of evaporative cooling}
\label{sec:evaporative}

Evaporative cooling can be used to reduce the temperature further. It seems
most efficient to apply the evaporation to the atoms, and sympathetically cool
the molecules, rather than to apply the evaporation directly to the molecules.
Therefore, we suppose that the evaporation is done in the magnetic trap by
applying an rf field which induces transitions between trapped and anti-trapped
Zeeman states at a value of magnetic field only reachable by the most energetic
atoms (an ``rf knife''). We study the sympathetic cooling of CaF when this
evaporative cooling is applied to Rb, for the two cases $a=+1.5\bar{a}$ and
$a=-0.5\bar{a}$. As the molecules cool, the molecular cloud shrinks: by
choosing an appropriate evaporative cooling ramp, the size of the atom cloud
can be optimized throughout the sympathetic cooling process.

We follow the theory and notation of evaporative cooling detailed in
\cite{Ketterle(1)96}. For simplicity, we assume that the atoms are held in a
harmonic trap. The rf knife is set so that an atom is lost if its energy
exceeds $\eta k_{\text{B}} T$, where $\eta$ is set quite large so that only the
high-energy tail of the distribution is cut off. The rate of change of atom
number $N_{\text{at}}$ follows
\begin{equation}\label{evap1}
\frac{d N_{\text{at}}}{d t} = -\frac{N_{\text{at}}}{\tau_{\text{ev}}},
\end{equation}
where $1/\tau_{\text{ev}}$ is the evaporation rate. It is given by
\begin{equation}\label{tauevap}
\tau_{\text{ev}} = \frac{\sqrt{2} e^{\eta}}{\eta} \tau_{\text{el}},
\end{equation}
where $\tau_{\text{el}}$ is the mean time between atom-atom elastic collisions
at the trap center. This scales with atom number as
\begin{equation}\label{collisionTimeScaling}
\frac{\tau_{\text{el}}}{\tau_{\text{el,i}}} =
\left(\frac{N_{\text{at}}}{N_{\text{at,i}}}\right)^{\alpha - 1},
\end{equation}
where $\alpha = \eta/3 - 1$ and the subscript i denotes the initial value.
Using Eqs.\,(\ref{evap1}), (\ref{tauevap}) and (\ref{collisionTimeScaling}), we
obtain
\begin{equation}\label{evap2}
\frac{1}{N_{\text{at,i}}} \frac{d N_{\text{at}}}{d t} =
-\frac{\kappa}{\tau_{\text{el,i}}}\left(\frac{N_{\text{at}}}{N_{\text{at,i}}}\right)^{2-\alpha},
\end{equation}
where $\kappa = \eta/(\sqrt{2} e^{\eta})$. The solution to this equation is
\begin{equation}
\frac{N_{\text{at}}(t)}{N_{\text{at,i}}} =
\left(1-\left(\alpha-1\right)\kappa\frac{t}{\tau_{\text{el,i}}}\right)^{1/(\alpha - 1)}.
\end{equation}

The mean time between collisions at the start of evaporation is
$\tau_{\text{el,i}} = 1/(\rho_0 \sigma \sqrt{2}\bar{v})=70.5$\,ms, where
$\rho_{0} = 10^{11}$\,cm$^{-3}$ is the initial density at the trap center,
$\sigma = 8\pi \times (95 a_{0})^{2}$ is the elastic cross section of $^{87}$Rb
at low temperature~\cite{Egorov(1)13}, and $\sqrt{2}\bar{v} = 0.22$\,m/s is the
mean relative velocity between two atoms at the initial temperature of
100\,$\mu$K. The temperature of the atoms scales as
$T_{\text{at}}/T_{\text{at,i}} = (N_{\text{at}}/N_{\text{at,i}})^{\alpha}$,
while the density scales as $n_{\text{at}}/n_{\text{at,i}} =
(N_{\text{at}}/N_{\text{at,i}})^{1 -3\alpha/2}$. In our simulations, we change
the atom number, temperature, density and radius in time, according to these
results. Otherwise, the simulation is unchanged. We stop the evaporation when
the atoms reach 1\,$\mu$K.

\begin{figure}[tb]
 \centering
 \includegraphics[width=0.47\textwidth]{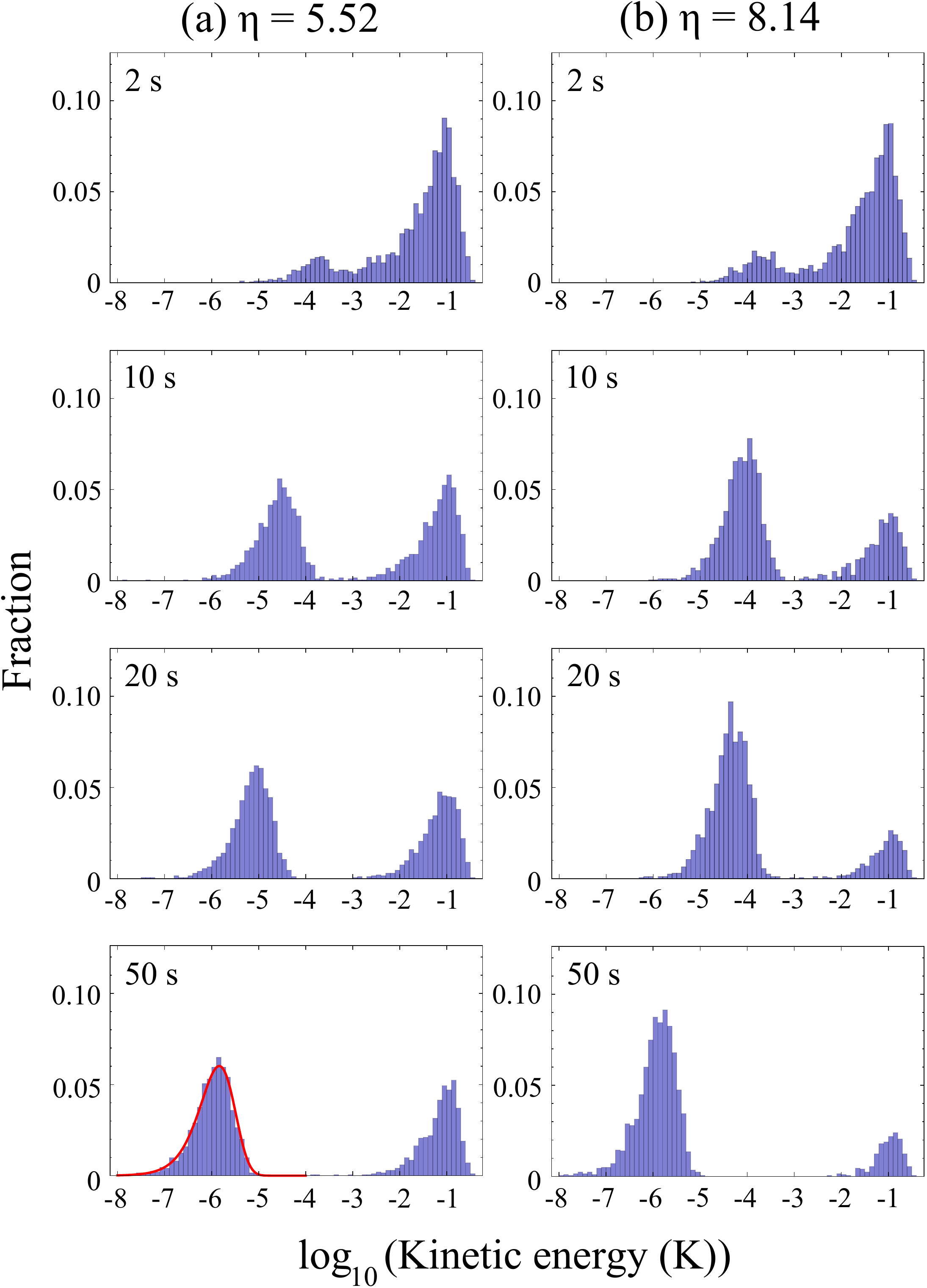}
\caption{\label{evhistogram} (Color online) Kinetic energy distributions at
four different times (2\,s, 10\,s, 20\,s and 50\,s) and for two values of the
evaporative cooling parameter: (a) $\eta = 5.52$, (b) $\eta = 8.14$. For
comparison, a Maxwell-Boltzmann distribution at 1\,$\mu$K is shown by a red
line. The coolant is Rb and $a=+1.5\bar{a}$.}
\end{figure}

Figure~\ref{evhistogram}(a) shows how the kinetic energy distribution of the
molecules evolves with time when $\eta=5.52$ and $a=+1.5\bar{a}$. At 2\,s, the
distribution is similar to the case without evaporation (see
Fig.~\ref{rbhistogram}(a)), but by 10\,s there is a large difference. For this
value of $\eta$ the atoms initially cool quickly, many atoms are ejected, and
the density gradually increases. About half the molecules cool along with the
atoms and these have kinetic energy below 100\,$\mu$K at 10\,s. The other half
remain uncooled because they find themselves outside the rapidly shrinking atom
cloud. After 50\,s the cold fraction is fully thermalized to the 1\,$\mu$K
temperature of the atom cloud. Figure~\ref{evhistogram}(b) shows the
corresponding evolution when $\eta =8.14$. In this case, the evaporation
initially proceeds slowly, and the molecule distribution remains similar to the
case without evaporation for the first 10\,s. Because the atom cloud shrinks
more slowly a larger number of molecules are captured into the cold fraction,
and these then cool to 1\,$\mu$K on a 50\,s timescale.

\begin{figure}[tb]
 \centering
 \includegraphics[width=0.45\textwidth]{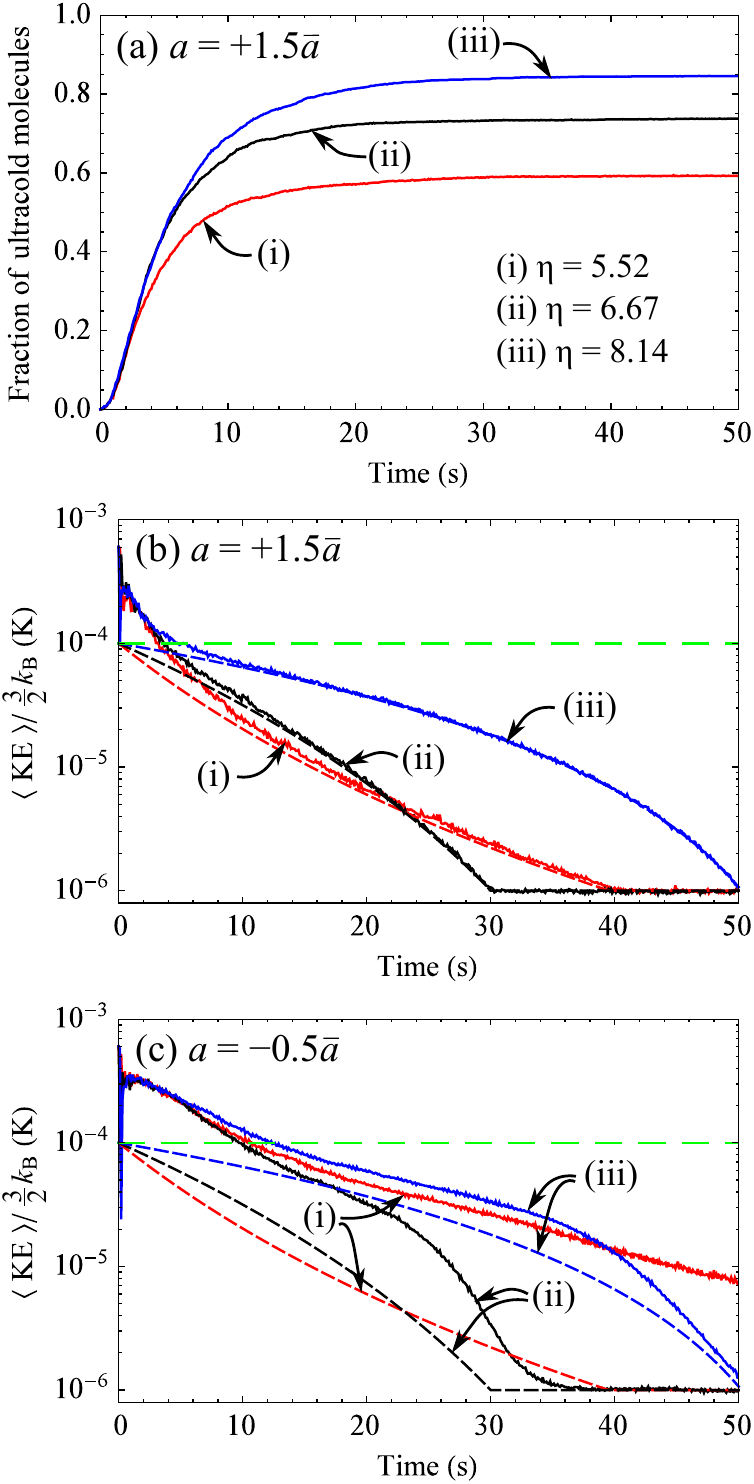}
\caption{\label{evaporative} (Color online) Sympathetic cooling of molecules
with evaporative cooling applied to the atoms. Graphs show the time evolution
of (a) the fraction of molecules with kinetic energy below 1\,mK, when $a=
+1.5\bar{a}$; (b) the mean kinetic energy of the ultracold fraction when $a=
+1.5\bar{a}$; (c) the mean kinetic energy of the ultracold fraction when $a=
-0.5\bar{a}$. (i, black) $\eta = 5.52$, (ii, red) $\eta = 6.67$, (iii, blue)
$\eta = 8.14$. In (b) and (c), the dashed lines show how the atomic temperature
evolves. The long-dash green line shows the atom temperature without
evaporative cooling.}
\end{figure}

Figure~\ref{evaporative}(a,b) show the fraction of molecules with kinetic
energy below 1\,mK, and the mean kinetic energy of that fraction, using
$a=+1.5\bar{a}$ and three different values of $\eta$: 5.52, 6.67, and 8.14.
When $\eta=8.14$ the atom cloud cools slowly at early times, and this gives the
molecules enough time to thermalize with the atoms before the atom cloud
shrinks too much. After this initial thermalization to the atom temperature,
the molecular temperature follows the evaporative cooling of the atoms very
closely. The ultracold fraction is high in this case, reaching 85\% after
50\,s. However, it takes the full 50\,s for this fraction to reach 1\,$\mu$K.
For this value of $\eta$, the atom density increases by a factor of 70 over
50\,s, and the mean atom-molecule collision rate increases from 4\,s$^{-1}$ to
45\,s$^{-1}$. When $\eta = 6.67$ the atoms cool more rapidly and the cloud size
shrinks more rapidly. Consequently, the ultracold fraction of molecules is
reduced to 74\% but this fraction now reaches 1\,$\mu$K in 30\,s. When $\eta =
5.52$ the atoms initially cool quickly, but the cooling rate slows down as time
goes on because the density does not increase rapidly enough to compensate for
the decrease in atom velocity. The ultracold fraction of molecules reduces to
59\%. The mean kinetic energy of this fraction falls quickly, reaching
100\,$\mu$K in 3.4\,s, and 10\,$\mu$K in 17\,s. Therefore, evaporative cooling
with a relatively low $\eta$ is a good strategy for cooling rapidly to
temperatures above 10\,$\mu$K. However, the cooling slows down at longer times
and it ultimately takes longer to reach 1\,$\mu$K than for the intermediate
value of $\eta$.

Finally, we consider the case where $a=-0.5\bar{a}$. This is a highly
unfavorable case compared to $a=+1.5\bar{a}$, both because the elastic cross
section in the ultracold limit is nine times smaller and because there is a
deep Ramsauer-Townsend minimum in the cross section for collision energies
slightly below 100\,$\mu$K, as can be seen in Fig.\,\ref{crosssection}. We find
that the fraction of molecules with kinetic energy below 1\,mK is almost
unchanged from that shown in Fig.\,\ref{evaporative}(a). This is to be expected
since, at energies higher than 1\,mK, the cross sections for the two values of
$a$ are not too different. Figure \ref{evaporative}(c) shows how the mean
kinetic energy of the ultracold fraction evolves when $a=-0.5\bar{a}$. Because
of the lower collision rate, the mean kinetic energy of the molecules lags
behind that of the atoms, instead of the two being locked together as they are
in the case of $a=+1.5\bar{a}$. The molecules are slow to reach 20\,$\mu$K for
all values of $\eta$, because they have to cool through the Ramsauer-Townsend
minimum to do so. For $\eta=5.52$, the atoms cool too quickly and the molecules
have not thermalized with the atoms even after 50\,s. For $\eta=8.14$ the
initial cooling rate of the atoms is slow enough that the molecule temperature
can more closely follow the atom temperature, both reaching 1\,$\mu$K in about
50\,s. The cooling of the molecules is fastest for the intermediate value of
$\eta$. In particular, the mean kinetic energy of the molecules falls rapidly
as soon as it is below 20\,$\mu$K, and it reaches 1\,$\mu$K in 36\,s. We see
that, even for this unfavorable value of $a$, evaporative cooling of the atoms
can bring the molecule temperature down to 1\,$\mu$K on a reasonable timescale,
provided a suitable value of $\eta$ is chosen. It is clear that knowledge of
the actual atom-molecule scattering length will be needed to choose the optimum
conditions for evaporative cooling.

\section{Conclusions}
\label{sec:conclusion}

In this paper, we have addressed the methodology for modeling sympathetic
cooling of molecules by ultracold atoms, and we have studied in detail the
results of simulations for a prototype case where ground-state CaF molecules in
a microwave trap are overlapped with ultracold Li or Rb atoms in a magnetic
trap. This work leads to a number of conclusions which we now summarize.

Previous work on sympathetic cooling used a hard-sphere model of
collisions based on an elastic cross section. This is appropriate at
very low energies (in the s-wave regime), but breaks down badly for heavy
molecules in the millikelvin regime. We have shown that a hard-sphere
model based on an elastic cross section significantly over-estimates the cooling rate for collision energies
above the s-wave scattering regime. A hard-sphere collision model that uses the
energy-dependent momentum transport cross section, $\sigma_{\eta}^{(1)}$, gives
the correct molecule cooling rate, but underestimates both the heating of the
atoms and the loss of atoms from the trap. We have therefore used the full
differential cross section to model atom-molecule collisions, so that the
cooling of the molecules and the associated heating and loss of atoms are all
modelled accurately.

We have studied sympathetic cooling of CaF with both Rb and Li over a range of
typical values of the atom-molecule scattering length $a$. We find that Rb
offers significant advantages over Li as a coolant for ground-state molecules.
The mean scattering length ${\bar a}$ is almost twice as large for Rb, and so
it is likely that the true scattering length will also be larger for Rb. The
mean energy transfer is proportional to $\mu/(m_{{\rm CaF}}+m_{{\rm at}})$
which is 0.48 for Rb, but only 0.19 for Li. If $a$ happens to be negative there
can be a deep Ramsauer-Townsend minimum in the cross section. For Li, the
minimum typically occurs when $E^{{\rm lab}}_{{\rm CaF}}$ is between 1 and
10\,mK, and the molecules cool very slowly because their energies must pass
through this minimum. For Rb, the minimum is shifted down an order of magnitude
in energy, and so the molecules do not encounter the minimum until they have
reached the ultracold regime. For Li, the cooling rate is very sensitive to the
actual value of $a$, while for Rb the initial cooling rate is fairly
insensitive to $a$ because the Rb+CaF cross section conforms closely to a
classical result, independent of $a$, down to temperatures near 1\,mK. This
brings less uncertainty about the likely results of sympathetic cooling
experiments if Rb is used. These advantages of Rb as a coolant are likely to
extend to other molecules of a similar or greater mass. Finally, it is
experimentally easier to prepare large, dense samples of ultracold Rb than of
ultracold Li.

It should be noted that the preference for Rb over Li applies
only to ground-state molecules that cannot be lost from the trap through
inelastic collisions. For molecules in static magnetic or electric traps, a
light collision partner such as Li, Mg or H provides a higher centrifugal
barrier than a heavy one such as Rb, and this may be important for suppressing
low-energy inelastic collisions \cite{Wallis:MgNH:2009, Wallis:LiNH:2011,
Gonzalez-Martinez:H+mol:2013}.

For molecules with an initial temperature of 70\,mK, cooled by
Rb with a temperature of 100\,$\mu$K and a peak density $10^{11}$\,cm$^{-3}$,
we find that, after 10\,s, 75\% of the molecules have cooled into a
distribution with a temperature of 200\,$\mu$K. If the initial temperature of
the molecules is reduced to 20\,mK, this fraction increases to 99\% due to
improved overlap between molecule and atom clouds. By arranging for the atom
trap depth to be far below the initial molecule temperature, we can ensure that
the majority of the energy in the molecule cloud is removed by atoms that are
lost from the trap, instead of heating the atom cloud. For efficient cooling
the atom number should exceed the molecule number by at least a factor of 100.
By applying evaporative cooling to the atoms, the molecules can be
sympathetically cooled more rapidly, or they can be cooled to far lower
temperatures. For values of the scattering length in the likely range, and with
a suitable choice of evaporation ramp, 70\% of the molecules can be cooled to
1\,$\mu$K within about 30\,s. These are all encouraging results: using
experimentally achievable atom numbers, densities and temperatures, sympathetic
cooling to ultracold temperatures can work on a timescale that is short
compared to achievable trap lifetimes. A good starting point for such
experiments would be a mixed-species magneto-optical trap of molecules and
atoms.

Data underlying this article can be accessed at http://dx.doi.org/10.5281/zenodo.32993 and used under the Creative Commons CCZero licence.

\acknowledgements This work has been supported by the UK Engineering and
Physical Sciences Research Council (grant EP/I012044/1 and EP/M027716/1), and by the European
Research Council.

\bibliographystyle{apsrev4-1}
\bibliography{../../all,extra}

\end{document}